\newcommand{\CLASSINPUTtoptextmargin}
{0.78in}

\newcommand{\CLASSINPUTbottomtextmargin}
{1.02in}

\documentclass[conference]{IEEEtran}
\makeatletter
\def\ps@headings{%
\def\@oddhead{\mbox{}\scriptsize\rightmark \hfil \thepage}%
\def\@evenhead{\scriptsize\thepage \hfil \leftmark\mbox{}}%
\def\@oddfoot{}%
\def\@evenfoot{}}
\makeatother 

\usepackage{cite}      
\usepackage{graphicx}  
\usepackage{psfrag}    
\usepackage{subfigure} 
\usepackage{url}       
\usepackage{stfloats}  
\usepackage{algorithm}
\usepackage{amsmath}   

\usepackage{array}
\usepackage{amssymb}
\usepackage{amsmath,amsfonts,amssymb,graphicx}
\usepackage{bm}
\usepackage{mathrsfs}

\usepackage{dsfont}

\usepackage{multicol}

\usepackage[end]{algpseudocode}

\usepackage{multirow}

\newtheorem{theorem}{\textbf{Theorem}}
\newtheorem{lemma}{\textbf{Lemma}}

\newtheorem{definition}{Definition}

\newcommand {\xli} {\lambda_i}
\newcommand {\xB} {\lambda_{-i}}
\newcommand {\xopt} {\lambda^*}
\newcommand {\xa} {\alpha}
\newcommand {\xlie} {\lambda_i^e}
\newcommand {\xBe} {\lambda_{-i}^e}

\IEEEoverridecommandlockouts

\begin{document}
\title{A Packet Dropping Mechanism for Efficient Operation of M/M/1 Queues with Selfish Users\thanks{This research was sponsored in part by the U.S. Army Research
Laboratory under the Network Science Collaborative Technology
Alliance, Agreement Number W911NF-09-2-0053, and by the U.S.
National Science Foundation under CNS-0831545. This work is an
extended version of the conference
paper~\cite{GaiLiuKrishnamachari:2011}.}}



\author{\IEEEauthorblockN{Yi Gai, Hua Liu, and Bhaskar Krishnamachari \\}
\IEEEauthorblockA{ Department of Electrical Engineering\\ University
of Southern California,
Los Angeles, CA 90089, USA\\
Email: $\{$ygai, hual, bkrishna$\}$@usc.edu} }


\maketitle

\begin{abstract}

We consider a fundamental game theoretic problem concerning selfish
users contributing packets to an M/M/1 queue. In this game, each
user controls its own input rate so as to optimize a desired
tradeoff between throughput and delay. We first show that the
original game has an inefficient Nash Equilibrium (NE), with a Price
of Anarchy (PoA) that scales linearly or worse in the number of
users. In order to improve the outcome efficiency, we propose an
easily implementable mechanism design whereby the server randomly
drops packets with a probability that is a function of the total
arrival rate. We show that this results in a modified M/M/1 queueing
game that is an ordinal potential game with at least one NE. In
particular, for a linear packet dropping function, which is similar
to the Random Early Detection (RED) algorithm used in Internet
Congestion Control, we prove that there is a unique NE. We also show
that the simple best response dynamic converges to this unique
equilibrium. Finally, for this scheme, we prove that the social
welfare (expressed either as the summation of utilities of all
players, or as the summation of the logarithm of utilities of all
players) at the equilibrium point can be arbitrarily close to the
social welfare at the global optimal point, i.e. the PoA can be made
arbitrarily close to 1. We also study the impact of arrival rate
estimation error on the PoA through simulations.
\end{abstract}


%
\IEEEpeerreviewmaketitle

\section{Introduction}\label{sec:intro}

In the past twenty years, the usage of the Internet has transitioned
from being primarily academic/research-oriented to one that is
primarily commercial in nature. In the current Internet environment,
each commercial entity is inherently interested only in its own
profit. Developing network mechanisms that are designed to handle
selfish behavior has therefore gained increasing attention in recent
years. The game theoretic approach, which was originally designed to
model and guide decisions in economic markets, provides a valuable
set of tools for dealing with selfish behavior
\cite{Srivastava:2005, Ozdaglar:2007, Altman:2006, Saad:2009,
Osborne:2004, MacKenzie:2006}.

In this work, we consider the network congestion problem at a single
intermediate store-and-forwarding spot in the network. Several users
send their packets to a single server with Poisson arrival rate. The
server processes the packets on a first come first serve (FCFS)
basis with an exponentially distributed service time. This is an
M/M/1 queueing model~\cite{Ross:1997}. There exists a trade-off in
this M/M/1 queueing model between throughput (representing the
benefit from service), and delay (representing the waiting cost in
the queue). In the gateway congestion control context
\cite{RFC:1994}, a measure that is widely used to describe this
trade-off is called ``Power", which is defined as the weighted ratio
of the throughput to the delay. When the users are selfish, we can
formulate a basic M/M/1 queueing game. In this game, we assume that
the users are selfish, and each control their own input arrival rate
to the server. Each user's utility is modeled to be the power ratio
for that user's packets.

This classic M/M/1 queueing game has been formulated and studied in
\cite{Kumar:1981,Douligeris:1992,Zhang:1992,Dutta:2003,Yi:2009}. The
results from these prior works and our own results in this work are
in agreement that the basic M/M/1 queuing game has an inefficient
Nash Equilibrium. We are therefore motivated to design an incentive
mechanism to force the users to operate at an equilibrium that is
globally efficient. In particular, we focus on the design of a
packet dropping scheme implemented at the server for this purpose.
Our objective is that the dropping scheme should be as simple as
possible, and it should minimize the Price of Anarchy (PoA, the
ratio of the social optimum welfare to the welfare of the worst Nash
equilibrium) to be as close to 1 as possible.

A key contribution of this work is the formulation of a modified
M/M/1 queuing game with a randomized packet dropping policy at the
server. We consider a simple and low overhead policy in our
formulation, wherein the server need only monitor the sum of the
rates of all users in the system. We show that this modified game
with a packet dropping scheme is an \emph{ordinal potential
game}~\cite{Monderer:1996}, which implies the existence of at least
one pure Nash Equilibrium.

We show first that utilizing a step-function for packet dropping
whereby the server drops all the packets when the sum-rate is
greater than a threshold (and none when the sum-rate is below the
threshold), results in infinite number of undesired Nash Equilibria
which harms the PoA.

This raises the question whether a more sophisticated approach can
do better. We show that indeed this is possible. In particular, we
develop an incentive mechanism with a linear packet dropping that
can improve the system efficiency to be arbitrarily close to the
global optimal point (i.e., a PoA arbitrarily close to 1). This
mechanism is similar to the Random Early Detection (RED) used for
congestion avoidance on the Internet~\cite{FloydJacobson:1993}. We
prove the uniqueness of NE of the game with this mechanism. We also
show that best response dynamics will converge to the unique NE.

Our paper is organized as follows. Section \ref{sec:related}
summarizes the related work. We present the model of an M/M/1 queue
game in Section \ref{sec:prob}. The social welfare and Price of
Anarchy are described in section \ref{sec:anarchy} to investigate
the efficiency of the NE. Then, in section \ref{sec:scheme}, we
propose to design an incentive packet dropping scheme implemented at
the server to improve the efficiency. Section \ref{sec:potential}
proves that the game defined with packet dropping policy is an
ordinal potential game by giving the potential function. Section
\ref{sec:best} shows the best response function. In section
\ref{sec:step}, we show the behavior when utilizing a simple
step-function for packet dropping. In section \ref{sec:linear} we
propose the RED-like linear packet dropping incentive scheme. We
show that with this scheme, it is possible to make the Price of
Anarchy arbitrarily close to the optimal point. The uniqueness of NE
of such a game is proved in section \ref{sec:uniqueness}. In section
\ref{sec:convergence}, we show that the best response dynamics will
converge to the unique Nash Equilibrium. In section
\ref{sec:estimation}, we undertake simulations to see how the
process of statistically estimating the input arrival rates in a
real system would impact the PoA. We conclude the work in section
\ref{sec:conclusion}.


\section{Related Work}\label{sec:related}

Throughput-delay tradeoffs in M/M/1 queues with selfish users have
been previously studied in
\cite{Kumar:1981,Douligeris:1992,Zhang:1992,Dutta:2003,Yi:2009}. A
utility function for each user is defined as the corresponding
application's power and each user is treated as a player in such a
game and adjusts its arrival rate to handle the trade-off between
throughput and delay. Every user is assumed to be selfish and only
wants to maximize its own utility function in a distributed manner.

Bharath-Kumar and Jaffe \cite{Kumar:1981} wrote one of the earliest
papers on the formulation of throughput-delay tradeoffs in M/M/1
queues with selfish users. The paper discusses the properties of
power as a network performance objective function. A class of greedy
algorithms where each user updates its sending rate synchronously to
the best response of all other users' rates to maximize the power is
proposed. Convergence of the best response to an equilibrium point
is shown in this paper.

Douligeris and Mazumdar \cite{Douligeris:1992} extended
Bharath-Kumar and Jaffe's work to the case with different weighting
factors defined in the power function for different users and
provided analytical results describing the Nash Equilibrium. They
showed that the equilibrium point that the greedy best response
dynamic algorithm converged to was a unique Nash Equilibrium.

The work by Zhang and Douligeris \cite{Zhang:1992} proved the
convergence of the best response dynamics for this basic M/M/1
queueing game under the multiple users case. Thus all these prior
works (\cite{Kumar:1981, Douligeris:1992, Zhang:1992}) studied only
variants of the basic game. Their work, along with ours, shows that
this basic game results in an inefficient outcome. Our work is the
first to develop a mechanism design for this problem that addresses
this shortcoming by showing how to achieve near-optimal performance
using a packet-dropping policy.

Dutta \textit{et al.} \cite{Dutta:2003} studied a related problem
involving a server that employs an oblivious active queue management
scheme, i.e. drops packets depending on the total queue occupancy
with the same probability regardless of which flow they come from.
They also consider an M/M/1 setting with users offering Poisson
traffic to a server with exponential service time. The users'
actions are the input rates and the utilities the goodput/output
rates. The existence and the quality of symmetric Nash equilibria
are studied for different packet dropping policies. Although our
work also explores oblivious packet dropping schemes, it is
different from and somewhat more challenging to analyze than
\cite{Dutta:2003}, because our utility function reflects the
tradeoff between goodput and delay.

In another, more recent work, \cite{Yi:2009}, Su and van der Schaar
have discussed linearly coupled communication games in which users'
utilities are linearly impacted by their competitors' actions. An
M/M/1 FCFS queuing game with the power as the utility function is
one illustrative example of linearly coupled communication games.
They also quantify the Price of Anarchy in this case, and
investigate an alternative solution concept called Conjectural
Equilibrium, which requires users to maintain and operate upon
additional beliefs about competitors.

There have been also several other papers related to queueing games,
albeit with different formulations. Haviv and
Roughgarden~\cite{Haviv:2007} considered a system with multiple
servers with heterogeneous service rates. Arrivals from customers
are routed to one of the servers, and the routing decisions are
analyzed based on NE or social optimization schemes. PoA is shown to
be upper bounded by the number of servers for the social optimum. Wu
and Starobinski \cite{Wu:2006} analyzed the PoA of $N$ parallel
links where the delays of links are characterized using unbounded
delay functions such as M/M/1 or M/G/1 queueing functions.
Economides and Silvester \cite{Economides:1991} studied a
multiserver two-class queueing game and developed the routing
policy.



For more general surveys on game theoretic formulations of
networking problems, we refer the reader to \cite{Altman:2006,
LiuKrishnamachariKapadia:2008}.

\section{Problem Formulation}\label{sec:prob}

\begin{figure}[h!]
  \centering
  \includegraphics[width=0.45\textwidth]{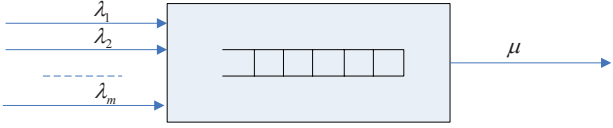}
 \caption{An M/M/1 queue} \label{fig:1}
\end{figure}

We consider a M/M/1 FCFS queue game as shown in Fig. \ref{fig:1}.
There are $m$ users with independent Poisson arrivals and the
arrival rates are $\lambda_1, \lambda_2, \dots, \lambda_m$. There is
a single server and the service time is exponentially distributed
with mean $\frac{1}{\mu}$.

We consider each user as a player for this game and the users are
selfish. Each player wants to maximize its own utility function by
adjusting its rate sending to the queue.

Note that there is a tradeoff between the throughput and delay for
each user, i.e., given the rates of all other users, if the input
rate increases, the delay increases too. In this paper, we consider
the measurement of this tradeoff between the throughput and delay of
the each user, and it is known as the ``power'', which is widely
used in the gateway congestion control context \cite{RFC:1994}. We
consider the power as the utility function of each user to measure
its throughput-delay tradeoff. For a given user $i$, the power is
defined as:
\begin{equation}
 \text{Power} = \frac{\text{Throughput}^{\alpha_i}}{\text{Delay}},
\end{equation}
where ${\alpha_i}$ is a parameter chosen based on the relative
emphasis placed on throughput versus delay. ${\alpha_i} > 1$ when
throughput is more important, while $0< {\alpha_i} < 1$ when we want
to emphasis delay more, and ${\alpha_i} = 1$ when the throughput and
delay are emphasized equally.

For M/M/1 queue, the throughput for user $i$ is $T_i = \xlie$ where
$\xlie$ is the effective rate served by the server. The delay for
user $i$ is calculated as: $D_i = \frac{1}{\mu - \sum\limits_{i =
1}^m \xlie}$.
So the power for user $i$ can be expressed as:
\begin{equation}
 P_i = \frac{T_i^{\alpha_i}}{D_i} = (\xlie)^{\alpha_i}(\mu - \sum\limits_{i = 1}^m \xlie).
\end{equation}

In this M/M/1 game, each player is selfish and wants to adjust its
arrival rate $\lambda_i$ to maximize its own utility function.
Throughout the paper, we assume that the queue is stable and thus $0
\leq \sum\limits_{i = 1}^m \xlie < \mu$.

When there is no dropping policy implemented at the server, $\xlie =
\xli$, and the optimization problem for each player $i$ is:
\begin{eqnarray}
 \max & & U_i(\lambda_i, \lambda_{-i}) =\lambda_i^{\alpha_i}(\mu - \sum\limits_{i = 1}^m \xli) \label{equ:4}\\
 s.t. & & \sum \lambda_i < \mu \nonumber\\
 & & \lambda_i \geq 0 \quad \forall i = 1, 2, \ldots, m \nonumber
  \end{eqnarray}

\section{Social Welfare and Price of Anarchy}\label{sec:anarchy}


In \cite{Douligeris:1992} the above M/M/1 queue game is studied and
a unique pure NE is proved to be:
\begin{equation}
\label{equ:7}
 \lambda_i^{NE} = \frac{\mu \alpha_i}{\sum\limits_{k = 1}^{m}\alpha_k + 1}, \forall \; i.
\end{equation}


When $\alpha_i = \alpha, \forall i$, this unique NE is expressed as
$\lambda_i^{NE} = \frac{\mu \alpha}{\alpha m + 1}, \forall \; i$.
Now suppose all users cooperate to achieve the maximal system
utility. We consider two ways to define the social optimal function:
the sum of the utilities of all the users and the sum of the
logarithm of the utilities of all the users. Defining the social
optimal function as the sum of the utilities of all the users is a
common way for evaluating the system efficiency and we present the
analysis results under this definition first. However, the fairness
among the users should also be considered and it is not revealed
under this definition; so we also consider a log-sum-utility social
welfare function which provides for utility fairness.

We can measure the efficiency of the system using two well known
measures called the Price of Anarchy (PoA) and Price of Stability
(PoS), that respectively compare the performance of selfish users in
the worst and best case Nash Equilibrium with the global optimum
achievable with non-selfish users. The definition of PoA and PoS of
a game $G$ is:
 \begin{equation}
 PoA(G) \triangleq \max\limits_{a \in \mathcal{E}(G)}
 \frac{U(a^{OPT})}{U(a)}.
\end{equation}
 \begin{equation}
 PoS(G) \triangleq \min\limits_{a \in \mathcal{E}(G)}
 \frac{U(a^{OPT})}{U(a)}.
\end{equation}
where $\mathcal{E}$ is the set of all the Nash Equilibriums in game
$G$.


\subsection{Sum-utility}

The optimization problem is defined as:
\begin{eqnarray}
 \max &\quad \sum \lambda_i^{\alpha} (\mu - \sum \lambda_i) \\
 s.t. &\quad 0 \leq \sum \lambda_i < \mu \nonumber\\
 &\quad \lambda_i \geq 0 \quad \forall i = 1, 2, \ldots, m \nonumber
\end{eqnarray}

Here we consider two cases:

1) $\alpha >1 $

First calculate $U^{OPT}$.
\begin{equation}
\begin{split}
 U^{OPT} & = \max_{\lambda_i} \sum \lambda_i^\alpha (\mu - \sum
 \lambda_i) \\
 & \leq \max_{\lambda_i} (\sum \lambda_i)^\alpha (\mu - \sum
\lambda_i)\\
 & = \max_{\lambda} \lambda^\alpha (\mu - \lambda)  \label{eqn:Uopt1}
  \end{split}
\end{equation}

We can get $\lambda^* = \frac{\mu \alpha}{\alpha + 1}$.

Equality holds in equation \ref{eqn:Uopt1} when $\lambda_i =
\lambda^*$ for some $i$, and $\lambda_i = 0, \forall j \neq i$.
Hence this is also the solution for $P_{sys}^{OPT}$.
\begin{equation}
 U^{OPT} = \frac{\alpha^\alpha \mu^{\alpha +
 1}}{(\alpha+1)^{\alpha+1}}. \nonumber
\end{equation}

Then we calculate $U^{NE}$ when players are selfish:
\begin{equation}
 U^{NE} = \frac{m \alpha^\alpha \mu^{\alpha + 1}}{(\alpha m
 +1)^{\alpha+1}}. \nonumber
\end{equation}

Note that there is only one NE in the game, so PoA and PoS are the
same and they are derived as below:
\begin{equation}
 PoA(G) = PoS(G) = \frac{U^{OPT}}{U^{NE}} =\frac{(\alpha m +1)^{\alpha+1}}{m (\alpha+1)^{\alpha+1}}.
\end{equation}

In this case we find that the PoA and PoS are proportional to
$m^\alpha$.

2) $\alpha <1 $

The calculation is similar as above, and details are omitted.
\begin{equation}
\begin{split}
 U^{OPT} & = \max_{\lambda_i} \sum \lambda_i^\alpha (\mu - \sum
 \lambda_i)\\ \nonumber
 & \leq m \max_{\lambda_i} (\frac{\sum \lambda_i}{m})^\alpha (\mu - \sum
\lambda_i)\\
 & =\max_{\lambda} m^{1-\alpha}\lambda^\alpha (\mu -
\lambda) .
\end{split}
\end{equation}

We also get $\lambda^* = \frac{\mu \alpha}{\alpha + 1}$,
\begin{equation}
 U^{OPT} = \frac{m^{1-\alpha} \alpha^\alpha \mu^{\alpha +
1}}{(\alpha+1)^{\alpha+1}} . \nonumber
\end{equation}

PoA and PoS are:
\begin{equation}
 PoA(G) = PoS(G) = \frac{U^{OPT}}{U^{NE}} =\frac{(\alpha m +1)^{\alpha+1}}{m^\alpha (\alpha+1)^{\alpha+1}}.
\end{equation}

In this case we find that the PoA and PoS are proportional to $m$.

Thus in both cases, we find that the PoA and PoS degrade linearly or
worse with the number of users.


\subsection{Sum-log-utility}

Now let's consider the sum of the logarithm of the utilities of all
the users. The reason we consider the logarithm function in the
social welfare is because when all users cooperate to achieve the
optimum, fairness among the users should also be considered, and a
logarithmic function would ensure this. The social welfare
optimization problem is:
\begin{eqnarray}
 \max & & \sum\limits_{i = 1}^m \log \left[ \lambda_i^{\alpha} (\mu - \sum\limits_{i = 1}^m \lambda_i) \right] \label{equ:2}\\
 s.t. & & 0 \leq \sum\limits_{i = 1}^m \lambda_i < \mu \nonumber\\
 & & \lambda_i \geq 0 \quad \forall i = 1, 2, \ldots, m  \nonumber
\end{eqnarray}

Note that for each player, maximizing the logarithm of its utility
function is equivalent to maximizing the utility function itself.
Therefore the NE remains the same as before.

Denote $\lambda = \sum\limits_{i = 1}^m \lambda_i$. We have the
following theorem for finding out the social optimum:

\theorem\label{claim:1} The solution for the social welfare
optimization problem is: $\xopt_i = \frac{\mu \xa}{m(\xa + 1)}$.


\begin{IEEEproof} see Appendix \ref{appendix:proof01}. \end{IEEEproof}

Note that $\lambda_i^{NE}$ is shown in (\ref{equ:7}) and by
substituting it into (\ref{equ:4}), we get the power for user $i$
as:
\begin{equation}
 U_{i}^{NE} = \frac{\alpha^\alpha \mu^{\alpha + 1}}{(\alpha m
 +1)^{\alpha+1}} \nonumber
\end{equation}

In general, the log-utility terms can be negative. To ensure that
both the numerator and denominator terms in the PoA and PoS are
non-negative in this case, we use a monotonic exponential mapping.
Note that there is only one NE in the game, so PoA and PoS are the
same, and they are derived as below:

\begin{equation}
\label{equ:8}
\begin{split}
  PoA(G) = PoS (G) & = \frac{e^{U_{}^{OPT}}}{e^{U_{}^{NE}}} =
  \frac{ \left( \frac{ \xa^\xa \mu^{\xa+1} }{ m^\xa (\xa + 1)^{\xa +1} }\right)^m }
  { \left( \frac{ \xa^\xa \mu^{\xa+1} }{ (\xa m + 1)^{\xa+1} } \right)^m }\\
  & =
  \left(\frac{ (\xa m + 1)^{\xa+1} }{ m^\xa (\xa + 1)^{\xa +1} }\right)^m > 1.
\end{split}
\end{equation}

From (\ref{equ:8}) we can see that PoA increases monotonically as
$m$ increases and goes to infinity as $m$ goes to infinity. So we
want to implement an incentive mechanism to improve the PoA.

\section{An Incentive Packet Dropping Scheme}\label{sec:scheme}

Note that $\lambda_i^{NE} = \frac{\mu \alpha}{\alpha m + 1} >
\frac{\mu \xa}{m(\xa + 1)} = \xopt_i$. This inspires us to find an
incentive packet dropping mechanism implemented at the server and we
wish this packet dropping mechanism to be as simple as possible. So
we consider the case where the server need only monitor the sum of
the rates of all users in the system and implement the packet
dropping policy based only on this information. Then the packet
dropping function could be expressed as $P_d(\sum \lambda_i)$. So
the optimization problem for each user $i$ with a dropping policy in
the system is:
\begin{equation}
\begin{split}
 \max &\qquad U_i(\lambda_i, \lambda_{-i}) =(\lambda_i (1 - P_d(\sum \lambda_i)))^{\alpha_i} \\
 & \qquad\qquad(\mu - \sum(\lambda_i(1 - P_d(\sum
 \lambda_i)))) \\
 s.t. &\qquad \sum \lambda_i (1 - P_d(\sum \lambda_i)) < \mu \\
 &\qquad  \lambda_i \geq 0 \quad \forall i = 1, 2, \ldots, m
\end{split}
\end{equation}

To facilitate the derivation, denote $P(\sum \lambda_i) = 1 -
P_d(\sum \lambda_i)$ and thus $P(\cdot)$ is the probability of
keeping packets in the system. Then the optimization problem for
each player $i$ is:
\begin{eqnarray}
 \max && U_i(\lambda_i, \lambda_{-i}) =(\lambda_i P(\sum \lambda_i))^{\alpha_i}\nonumber\\
 &&\qquad\qquad(\mu - \sum(\lambda_i P(\sum
 \lambda_i)))) \label{equ:30}\\
 s.t. && \sum \lambda_i P(\sum
 \lambda_i) < \mu \label{equ:31}\\
 && \lambda_i \geq 0 \quad \forall i = 1, 2, \ldots, m
\end{eqnarray}

 We denote the above game as $G_p = (N,
\{\mathcal{A}_i\}, \{U_i\} )$.

\section{Potential Game}\label{sec:potential}

In this section, we prove that when the dropping function is a
function which only depends on the sum of total incoming rates, the
game is a potential game and thus there exists at least one pure NE.

\definition a game $G = (N, \{\mathcal{A}_i\}, \{U_i\})$ is called
an \emph{ordinal potential game} if there exists a global function
$\phi : \mathcal{A} \longrightarrow \mathds{R}$ such that for every
player $i \in N$, for every $a_{-i} \in \mathcal{A}_{-i}$ and for
every $a'_i, a_i'' \in \mathcal{A}_i$,
\begin{equation}
 sgn(U_i(a_i', a_{-i}) - U_i(a_i'', a_{-i}) ) = sgn(\phi(a_i', a_{-i}) - \phi(a_i'', a_{-i}) )
\end{equation}
where $sgn(x)$ is the sign function that takes on the value -1 when
$x<0$, 0 when $x=0$, and 1 when $x > 0$. 

Also, the following Theorem \ref{theorem:1} holds for the existence
of NE in a potential game:

\theorem\label{theorem:1} [Monderer-Shapley,
1996~\cite{Monderer:1996}] Every potential game with finite-players,
continuous utilities, and compact strategy sets possesses at least
one pure-strategy equilibrium.

Now we will prove that the M/M/1 queueing game with a packet
dropping function $P_d(\sum \lambda_i)$ is a potential game.

\theorem\label{claim:potential} $G_p$ is a ordinal potential game
with
potential function \\
\scalebox{0.9}{ $
 \phi(\lambda_1, \lambda_2, \dots, \lambda_m) = \left(\mu - P(\sum \lambda_i) \sum\limits_{i = 1}^m \lambda_i
 \right) \left( \prod\limits_{i = 1}^m (\lambda_i P(\sum
 \lambda_i))^{\xa_i} \right)
$ }

\begin{IEEEproof}
\begin{equation}
 \begin{split}
 & \phi(\xli', \xB) - \phi(\xli'', \xB)\nonumber \\
 & = \left(\mu - P(\lambda_i' + \xB) ( \sum\limits_{j \neq i }^m
 \lambda_j + \xli')
 \right) \\
 &\qquad\qquad\left( \prod\limits_{j \neq i}^m (\lambda_j P(\sum
 \lambda_j))^{\xa_j} \right) \xli'^{\xa_i} P(\lambda_i' + \xB)^{\xa_i} \\
 & \quad - \left(\mu - P(\lambda_i'' + \xB) ( \sum\limits_{j \neq i }^m
 \lambda_j + \xli'')
 \right) \\
 &\qquad\qquad\left( \prod\limits_{j \neq i}^m (\lambda_j P(\sum
 \lambda_j))^{\xa_j} \right) \xli''^{\xa_i} P(\lambda_i'' + \xB)^{\xa_i} \\
 & = \left( \prod\limits_{j \neq i}^m (\lambda_i P(\sum
 \lambda_i))^{\xa_i} \right) \left[
 (\mu - P(\lambda_i' + \xB)\right.\\
 &\qquad\qquad( \sum\limits_{j \neq i }^m
 \lambda_j + \xli')
 ) (\xli' P(\lambda_i' + \xB))^{\xa_i}\\
 &  - (\mu - P(\lambda_i'' + \xB) ( \sum\limits_{j \neq i }^m
 \lambda_j + \xli'')
 ) (\xli'' P(\lambda_i'' + \xB))^{\xa_i}
 ]\\
 & = \left( \prod\limits_{j \neq i}^m (\lambda_i P(\sum
 \lambda_i))^{\xa_i} \right) \left(U_i(\xli', \xB) - U_i(\xli'', \xB) \right)
 \end{split}
\end{equation}
\end{IEEEproof}

Note that $G_p$ has a finite number of players and continuous
utilities. However its strategy sets are not compact in
(\ref{equ:31}) so we could not directly apply Theorem
\ref{theorem:1} to claim there exists at least one NE in $G_p$. But
we modify the $G_p$ to be the equivalent game as follows:
\begin{equation}
\begin{split}
 \max &\qquad U_i(\lambda_i, \lambda_{-i}) =(\lambda_i P(\sum \lambda_i)))^{\alpha_i}\\
 &\qquad\qquad  (\mu - \sum(\lambda_i P(\sum
 \lambda_i))))\\
 s.t. &\qquad  \sum \lambda_i P(\sum
 \lambda_i) \leq \mu \\
 &\qquad  \lambda_i \geq 0 \quad \forall i = 1, 2, \ldots, m
\end{split}\label{eqn:compactStrategy}
\end{equation}
since any solution to the maximization problem in $G_p$ will not
satisfy $\sum \lambda_i P(\sum
 \lambda_i) = \mu$. Now strategy sets of $G_p$ are compact and thus there exists at least one NE in $G_p$.

Note that in the following when we describe PoA and PoS for the
packet dropping game $G_p$, we respectively compare the worst and
best NE obtained for this game with respect to the social welfare
(global optimum) that can be obtained through cooperation without
packet dropping.

\section{Best Response Function}\label{sec:best}

From now, for tractability, we consider the case $\alpha_i = \alpha,
\forall i$ for our proposed incentive packet dropping scheme.

$\forall i$,  let
\begin{equation}
\frac{\partial U_i(\lambda_i, \lambda_{-i}')}{\partial \lambda_i}=
0. \nonumber
\end{equation}

If $P(\sum \lambda_i)$ is differentiable with respect to
$\lambda_i$, we will have
\begin{equation}
\label{equ:9}
\begin{split}
& \xa P \mu - (\xa +1) P P' \xli \xB - (\xa + 1) P P'
\xli^2 \\
& - \xa P^2 \xB - (\xa+1) P^2 \xli + \xa P' \xli \mu = 0
\end{split}
\end{equation}
where $P'$ is the derivative of $P(\sum \lambda_i)$ with respect to
$\lambda_i$.

The above defines an implicit best response function
$\mathfrak{F}(\xli, \xB) = 0$ which shows the relationship between
$\xli$ and $\xB$.

\section{Step Dropping Function}\label{sec:step}

\begin{figure}[h]
  \centering
  \includegraphics[width=0.2\textwidth]{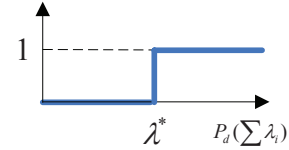}
 \caption{Step dropping function $P_d(\sum \lambda_i)$} \label{fig:2}
\end{figure}

An intuitive dropping policy that first comes to mind is a step
function as shown in Fig. \ref{fig:2}.

The expression of $P_d(\sum \lambda_i)$ is:
\begin{equation}
P_d(\sum \lambda_i) = \left\{
\begin{array}
 {r@{\quad:\quad}l}
 0 & \sum \lambda_i \leq \lambda^*\\
 1 & \sum \lambda_i > \lambda^*
\end{array}
\right.
\end{equation}

We have the following result for the corresponding packet dropping
game.

\theorem \label{claim:c3} $\lambda'$ is a N.E. if and only if $\sum
\lambda_i' = \lambda^*$.

\begin{IEEEproof} see Appendix \ref{appendix:proof02}. \end{IEEEproof}

Based on Theorem \ref{claim:c3}, the NEs of the game with the step
dropping function are not unique. $\text{PoS} = 1$ since there
exists a NE with $\xli = \xopt_i, \forall i$. However, in the
sum-utility case, $\text{PoA} = m^{\xa-1}$ when $\xa > 1$, and
$\text{PoA} = m^{1-\xa}$ when $\xa < 1$. Moreover, PoA is infinite
in the sum-log-utility case since there exists a NE which has one
user $i$ with $\xli = 0$. Hence this is not a desirable result for
improving the efficiency. We therefore next consider a slightly more
sophisticated dropping function that has a linear profile.

\section{Linear Dropping Function}\label{sec:linear}

\begin{figure}[h]
  \centering
  \includegraphics[width=0.45\textwidth]{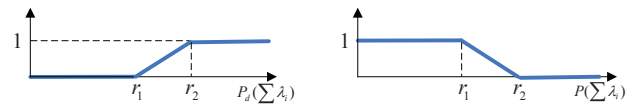}
 \caption{Illustration of $P_d(\sum \lambda_i)$ and $P(\sum \lambda_i)$} \label{fig:3}
\end{figure}

We consider the game with the following linear function of $P(\sum
\lambda_i)$ (and thus the packet dropping function $P_d(\sum
\lambda_i) = 1 - P(\sum \lambda_i)$ is also a linear function.) Fig.
\ref{fig:3} illustrates $P_d(\sum \lambda_i)$ and $P(\sum
\lambda_i)$.
\begin{equation}
\label{equ:15} P(\sum \lambda_i) = \left\{
\begin{array}
 {l@{\quad:\quad}l}
 1 & 0 \leq \sum \lambda_i \leq r_1\\
 A (\sum \lambda_i) + D & r_1 \leq \sum \lambda_i \leq r_2\\
 0 & \sum \lambda_i \geq r_2
\end{array}
\right.
\end{equation}
where
\begin{equation} \label{eq:1} \left\{ \begin{aligned}
         A & = \frac{1}{r_1 - r_2} \\ \nonumber
         D & = -A r_2
         \end{aligned} \right.
\end{equation}
\begin{equation}
 P' = \frac{\partial P}{\partial \xli } = A . \nonumber
\end{equation}

For linear dropping scheme, (\ref{equ:9}) becomes:
\begin{equation}
\label{equ:10}
\begin{split}
 & \xa P \mu - (\xa +1) P A \xli \xB - (\xa + 1) P A
\xli^2 \\
& - \xa P^2 \xB - (\xa+1) P^2 \xli + \xa A \xli \mu = 0
\end{split}
\end{equation}
The above also defines an implicit function $\mathfrak{F}(\xli, \xB)
= 0$.

Denote $\xlie = P(\sum \lambda_i)\xli$ and $\xBe = P(\sum
\lambda_i)\xB$. First we want find out if we could design a dropping
policy in this linear form such that the system could have a NE that
is the same as the social optimum. If not, we will then explore how
much efficiency it could achieve.

\theorem\label{claim:c4} There does not exist a linear packet
dropping policy such that $PoA=1$.

\begin{IEEEproof}

Assume the above Theorem does not hold, when $P \neq 0$,
substituting $\xli$ and $\xB$ with $\xlie/P$ and $\xBe/P$ we have,
\begin{equation}
\begin{split}
& \xa P \mu - \frac{1}{P}(\xa +1) A \xlie \xBe - \frac{1}{P}(\xa +
1) A (\xlie)^2 \\
 & \quad - \xa P \xBe - (\xa+1) P \xlie + \frac{1}{P}\xa A
\xlie
\mu = 0\\[1mm] \nonumber
 \Longrightarrow \quad & P (\xa \mu - \xa \xBe - (\xa
+1)\xlie) \\
& = \frac{1}{P} [ (\xa +1)A \xlie \xBe + (\xa +1)A (\xlie)^2 - \xa A
\xlie \mu ] \nonumber
  \end{split}
\end{equation}

Since $\xopt = \frac{\mu \xa}{\xa + 1}$ implies $\xa \mu  = (\xa +
1) \xopt$, we have
\begin{equation}
\label{equ:11}
\begin{split}
\quad& P [(\xa + 1) \xopt - \xa \xBe - (\xa +1)\xlie] \\
 \quad& =\frac{1}{P}  A \xlie[ (\xa +1)(\xBe + \xlie) - (\xa +1) \xopt
 ] \nonumber
\end{split}
\end{equation}
\begin{equation}
\label{equ:11}
 \begin{split}
 \Longrightarrow &\quad P [(\xa + 1) (\xopt - \xBe - \xlie) + \xBe] \qquad\qquad\\
 &= \frac{1}{P}  A \xlie (\xa +1)(\xBe + \xlie - \xopt )
\end{split}
\end{equation}

Note that $P\xli + P\xB = \xopt$ implies that $\xlie + \xBe =
\xopt$. So the right-hand side of the equality is $0$. While the
left-hand side of the equality is $P \xBe$. Since $\xBe \neq 0$, so
$P \xBe \neq 0$. Thus the left-hand side of the equality is not $0$
and this leads to a contradiction. Therefore Theorem \ref{claim:c4}
holds.
\end{IEEEproof}

Theorem \ref{claim:c4} shows that we could not design a linear
packet dropping policy with $PoA = 1$. The following theorem shows
that we could design an incentive packet dropping policy such that
PoA could be arbitrarily close to 1.

\theorem\label{claim:c5} Given any $\epsilon$, there exists a linear
packet dropping policy such that $1< PoA \leq 1+ \epsilon$.

\begin{IEEEproof}

Note (\ref{equ:11}) in the proof of Theorem \ref{claim:c4} implies:
\begin{equation}
\label{equ:12}
 P^2 [(\xa + 1) (\xopt - \xBe - \xlie) + \xBe] =
 A \xlie (\xa +1)(\xBe + \xlie - \xopt )
\end{equation}

The right-hand side of (\ref{equ:12}) is greater than $0$ only when
$\xBe + \xlie < \xopt$ (note that $A < 0$). Then given $A$ and
$\xlie, \xBe$ such that $\xBe + \xlie < \xopt$, we will have a
solution for $P^2$ and thus we could get the value of $D$.

This means that we can design a packet dropping scheme such that it
has a NE that satisfies $p\xli + p\xB \longrightarrow \xopt$ from
the left side (left approximation). If we can further prove that the
NE is unique in this game (see Theorem \ref{claim:c6}), then give
any $\epsilon
> 0$, we could find a linear packet dropping policy at the server
such that $1< PoA \leq 1+ \epsilon$.

\end{IEEEproof}


We propose Algorithm \ref{alg1} to show how to design the parameters
$r_1$ and $r_2$ in our proposed incentive packet dropping policy to
achieve a desired PoA such that $1< PoA \leq 1+ \epsilon$ given any
$\epsilon$. We denote $\lambda_e = \sum \lambda_i^e$. Line
\ref{line:1} ensures that the sum of the rates will be less than
$\mu$ before the server starts to drop packets. $\tilde{p}$ is the
value of $P(\sum \lambda)$ at the desired NE which is derived from
the desired PoA. $\tilde{p} = 1 - Pr\{\text{the packet dropping
probability at desired NE}\}$. Line \ref{line:2} is the calculation
of desired NE. The choice of $\widetilde{\lambda}$ is based on the
desired value of PoA, i.e., given $\epsilon
> 0$, we could accordingly derive the
value of a desired sum rate such that $1< PoA \leq 1+ \epsilon$.
Since $(\widetilde{\lambda_e}, \widetilde{p})$ is a solution of
$P(\sum \lambda)$, line \ref{line:3} shows how to therefore get the
expression of $A$ and $D$. Then at line \ref{line:4}, we could solve
the equation (\ref{equ:10}) given all the values above and get the
value of $r_2$. Based on the result of $r_2$, the value of $r_1$ is
calculated.

\begin{algorithm}

\caption{Parameter Calculation for Incentive Packet Dropping Scheme}
\label{alg1}

\textbf{Input:} PoA bound parameter $\epsilon$

\textbf{Output:} $r_1$ and $r_2$ of our proposed incentive packet
dropping policy in (\ref{equ:15}) such that $1< PoA \leq 1+
\epsilon$.

\begin{algorithmic}[1]
\State \label{line:1} Pick any $\tilde{p}$ such that $\frac{a}{a+1}<
\tilde{p} < 1$.

\State \label{line:2} Calculate a desired sum rate
$\widetilde{\lambda}$, of which
\begin{equation}
 \widetilde{\lambda} = \{ \frac{\widetilde{\lambda_e}}{m
\widetilde{p}},  \frac{(m-1)\widetilde{\lambda_e}}{m
\widetilde{p}}\}.
\end{equation}
is the desired NE such that $1< PoA \leq 1+ \epsilon$. Note that
$\lambda_e = \tilde{p} \lambda$.

\State \label{line:3}  Suppose $P(\sum \lambda_i) = A \sum \lambda_i
+ D$ pass through the point $(\widetilde{\lambda}, \widetilde{p})$.
Then we have
\begin{equation} \label{equ:16}
  A = \frac{\widetilde{p}}{\widetilde{\lambda} - r_2}, D = -\frac{\widetilde{p} r_2}{\widetilde{\lambda} -
  r_2}.
\end{equation}

\State \label{line:4} Insert the above values of the variables into
(\ref{equ:10}) and get the value of $r_2$.

\State \label{line:5} Insert the value of $r_2$ into (\ref{equ:16})
and get the value of $A$. Then $r_1 = \frac{1}{A} + r_2$
\end{algorithmic}
\end{algorithm}

Note that (\ref{equ:10}) is a quadratic equation for the parameters
$A$ and $D$ given the values of all the other variables. But with
Algorithm \ref{alg1}, we could always find a unique solution as
stated in Theorem \ref{claim:c5}.

\theorem \label{claim:c5} Algorithm \ref{alg1} yields a unique
linear packet dropping scheme, i.e., unique values for $r_1$ and
$r_2$ for any PoA bound.

\begin{figure}[h]
  \centering
  \includegraphics[width=0.30\textwidth]{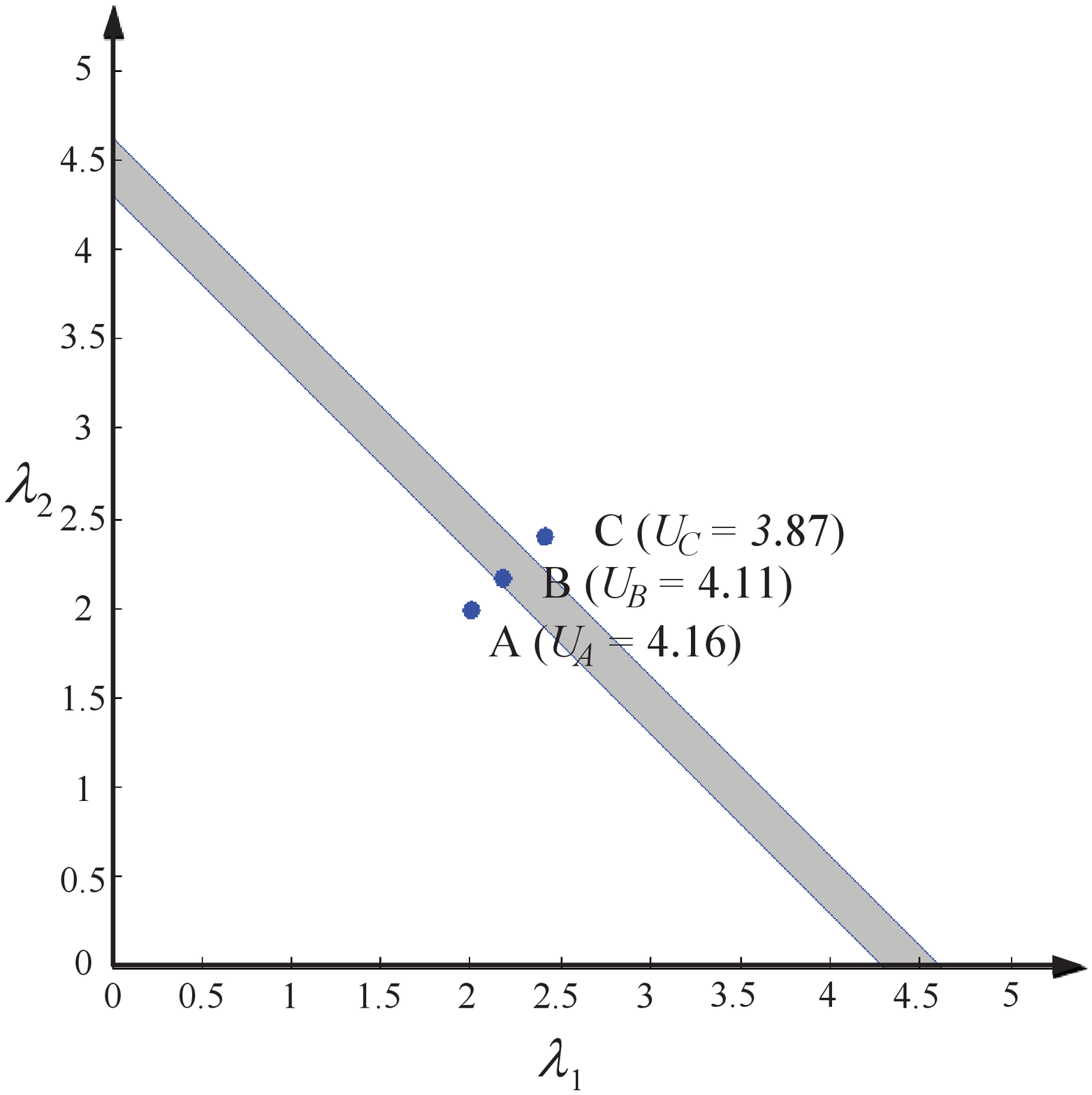}
 \caption{An example of our incentive packet dropping scheme.} \label{fig:6}
\end{figure}

\begin{IEEEproof}

After inserting the value $\widetilde{p}$, and $\widetilde{\lambda}
= \{ \frac{\widetilde{\lambda_e}}{m \widetilde{p}},
\frac{(m-1)\widetilde{\lambda_e}}{m \widetilde{p}}\}$ into
(\ref{equ:10}) at line \ref{line:4}, we have the equality
\begin{equation}
\label{equ:17}
 \begin{split}
 & \widetilde{p}^2 [(\xa + 1) (\xopt
  - \widetilde{\lambda_e}) +  \frac{(m-1)\widetilde{\lambda_e}}{m
\widetilde{p}}] \\
& = \frac{\widetilde{p}}{\widetilde{\lambda_e}/\widetilde{p} - r_2}
\frac{\widetilde{\lambda_e}}{m \widetilde{p}} (\xa
+1)(\widetilde{\lambda_e} - \xopt )
\end{split}
\end{equation}

It is obvious that the above is a linear equation of the variable
$r_2$. And thus we could get a unique solution of $r_2$. Therefore,
there is a always a unique solution of $r_1$ and $r_2$ provided by
Algorithm \ref{alg1}.
\end{IEEEproof}
Our proposed packet dropping scheme is similar to the Random Early
Detection (RED) algorithm. It is simple and easy to be implemented
with low overhead at the server. Fig. \ref{fig:6} shows an example
of our linear packet dropping policy with $\mu = 6$, $m = 2$ and
$\xa = 2$. (\ref{equ:2}) and (\ref{equ:8}) are used to calculate the
utility and PoA. Point A represents $\xopt$, which is then
calculated to be 4 ($\lambda_1 = \lambda_2 = 2$). We assume
$\tilde{p} = 0.9$, PoA bound parameter $\epsilon = 0.05$.
Implementing Algorithm \ref{alg1} with Matlab, we pick
$\widetilde{\lambda_e} = 3.9$, we then get $r_1 = 4.3012$, $r_2 =
4.622$, $A = -3.1154$, $D = 14.4000$. Point B represents the Nash
Equilibrium with our proposed packets dropping policy. Point C
represents the Nash Equilibrium of the original game without packet
dropping policy. The shaded area shows the cases where packet
dropping happens. For the comparison, the utilities are shown in the
figure and we can see that $PoA$ is improved from $1.3396$ to
$1.0455$.

\section{Uniqueness of NE}\label{sec:uniqueness}

If we use the packet dropping scheme in algorithm \ref{alg1} we are
guaranteed that the game $G_p$ always has a NE with the desired PoA
bound. Now our question is whether the scheme yields a unique NE.
This is important not only for finding out whether our proposed
scheme is efficient but also for the convergence issues. As surveyed
in \cite{Altman:2009}, there are not many general results on
equilibrium uniqueness. We were unable to find any existing theorem
that we could use directly to prove the uniqueness of NE in our
M/M/1 queueing game. This makes the analysis of this incentive
design problem more challenging.

\theorem \label{claim:c6} There is a unique NE for the M/M/1 Game
with the linear packet dropping scheme described in
Algorithm~\ref{alg1}.

To prove \ref{claim:c6}, we first prove the following three lemmas.

\lemma \label{lemma:1} $|A|$ increases monotonically as $(\xopt
-\widetilde{\lambda_e})$ decreases where $\widetilde{\lambda_e}$ is
the total rate of all users at desired NE (as in Algorithm
\ref{alg1}).

\begin{IEEEproof} Note that (\ref{equ:17}) is equivalent to:
\begin{equation}
\label{equ:21}
 \widetilde{p}^2 [(\xa + 1) (\xopt - \widetilde{\lambda_e}) +  \frac{(m-1)\widetilde{\lambda_e}}{m
\widetilde{p}}] =
 A \frac{\widetilde{\lambda_e}}{m
\widetilde{p}} (\xa +1)(\widetilde{\lambda_e} - \xopt ) \nonumber
\end{equation}
This means
\begin{equation}
\label{equ:20} \frac{(m-1)\widetilde{\lambda_e}}{m \widetilde{p}
(\xopt - \widetilde{\lambda_e})} = (\xa
+1)\left(|A|\frac{\widetilde{\lambda_e}}{m \widetilde{p}} -
\widetilde{p}^2 \right).
\end{equation}

Note that as $(\xopt -\widetilde{\lambda_e})$ decreases,
$\widetilde{\lambda_e}$ increases and $\frac{1}{\xopt -
\widetilde{\lambda_e}}$ increases, so the left-hand side of
(\ref{equ:20}) increase. This implies $|A|$ increases, and thus
Lemma \ref{lemma:1}.
\end{IEEEproof}

\lemma \label{lemma:2} $\forall \xB < r_1, \frac{\partial
U_i(\lambda_i, \lambda_{-i})}{\partial \lambda_i} > 0$ at $\lambda_i
= (r_1 - \xB)^+$.

\begin{IEEEproof}

Note that $P = 1$ at $r_1$, then
\begin{equation}
 \begin{split}
&\;\; \frac{\partial U_i(\lambda_i, \lambda_{-i})}{\partial \lambda_i} \nonumber\\
 = &\;\; \xa \mu - (\xa +1) A \xli \xB - (\xa + 1) A \xli^2 - \xa \xB \\
 &\;\; -
(\xa+1) \xli + \xa A \xli \mu \nonumber\\[0mm]
  = &\;\; \xa \mu - \xa \xB - (\xa +
1) \xli \\
&\;\; - A \xli [ (\xa + 1) (\xB +
\xli) - \xa \mu ]\nonumber\\[0mm]
  = &\;\; \xa \mu - (\xa + 1) (\xB + \xli)  + \xB \\
  &\;\; - A \xli [ (\xa + 1)
(\xB +
\xli) - \xa \mu ]\nonumber\\[0mm]
  = &\;\; [\xa \mu - (\xa + 1) (\xB + \xli) ](1+ A \xli) + \xB \nonumber\\[0mm]
  = &\;\; [\xa \mu - (\xa + 1) r_1 ](1+ A \xli) + \xB \nonumber\\[0mm]
  = &\;\; [(\xa + 1)\xopt - (\xa + 1) r_1 ](1+ A \xli) + \xB \nonumber\\[0mm]
   = &\;\; (\xa + 1)(\xopt - r_1 )(1+ A (r_1 - \xB)) + \xB \label{equ:22}
\end{split}
\end{equation}

Denote (\ref{equ:22}) as $g(\lambda_{-i})$. Then,
\begin{equation}
 \label{equ:23}
 \frac{\partial g}{\partial \lambda_{-i}} = -(\xa + 1)(\xopt - r_1 )A +
 1.
\end{equation}
When $|A|$ is large enough such that $r_1 > \xopt$ and $|A| >
\frac{1}{(\xa + 1)(r_1 -\xopt)}$, $\frac{\partial g}{\partial
\lambda_{-i}} < 0$.

From Lemma \ref{lemma:1} we know that as $\widetilde{\lambda_e}$
gets closer to $\xopt$, $|A|$ increases. This means that when we
design a dropping policy with $PoA$ approaching to $1$, $|A|$ could
be large enough such that $r_1 > \xopt$ and $|A| > \frac{1}{(\xa +
1)(r_1 -\xopt)}$.

So $g(\lambda_i)$ achieves the minimum value when $\xB = r_1$. Then
\begin{equation}
 \begin{split}
\frac{\partial U_i(\lambda_i, \lambda_{-i})}{\partial \lambda_i}
&\;\;>\;\;
(\xa + 1)(\xopt - r_1) + r_1\\
&\;\;=\;\; (\xa + 1)\xopt - \xa r_1\\
&\;\;=\;\; \xa \mu - \xa r_1 > 0.
\end{split}
\end{equation}
\end{IEEEproof}

\begin{figure}[h]
  \centering
  \includegraphics[width=0.3\textwidth]{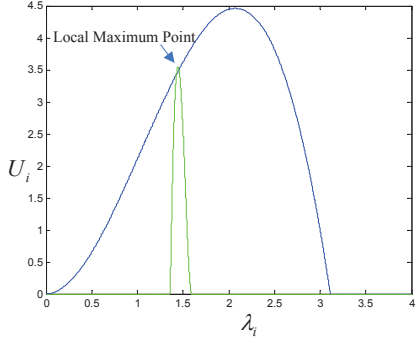}
 \caption{Illustration of local maximum point of $U_i(\lambda_i,
\lambda_{-i})$} \label{fig:4}
\end{figure}

Lemma \ref{lemma:2} implies that given $\xB$, $U_i(\lambda_i,
\lambda_{-i})$ which is a function of $\lambda_i$, has a maxima at
$r_1- \lambda_{-i} < \lambda_i  < r_2 - \lambda_{-i}$ as shown in
Fig. \ref{fig:4}.

\begin{figure}[h]
  \centering
  \includegraphics[width=0.3\textwidth]{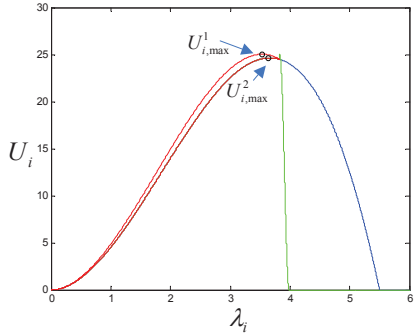}
 \caption{Illustration of local maximum point of $U_{i, max}^1$ and $U_{i, max}^2$} \label{fig:5}
\end{figure}


Given $\xB$, if $\xB + \frac{(\mu - \xB)\xa}{\xa + 1} < r_1$, then
$U_i(\lambda_i, \lambda_{-i})$ will reach a local maximal point when
$\lambda_i = \frac{(\mu - \xB)\xa}{\xa + 1}$ as shown in Fig.
\ref{fig:5}. Note that $P(\sum \lambda_i) = 1$ at this point. We
denote this local maximal value as $U_{i, max}^1$. Then
$U_i(\lambda_i, \lambda_{-i})$ will reach another local maximal
point with $\lambda_i^e = \frac{(\mu - \xB)\xa}{\xa + 1}$. Note that
$P(\sum \lambda_i) < 1$. We denote this local maximal value as
$U_{i, max}^2$.

\lemma \label{lemma:3} Given $\xB$, if $\xB + \frac{(\mu -
\xB)\xa}{\xa + 1} < r_1$, $U_{i, max}^1 < U_{i, max}^2$.

\begin{IEEEproof}

Note that
\begin{equation}
U_{i, max}^1 =(\lambda_i^1)^{\alpha}(\mu - (\lambda_i^1) - \xB),
\nonumber
\end{equation}
where $\lambda_i^1 = \frac{(\mu - \xB)\xa}{\xa + 1}$.
\begin{equation}
U_{i, max}^2 = (P \lambda_i^2)^{\alpha}(\mu - (P \lambda_i^1) - P
\xB). \nonumber
\end{equation}
Denote $\xlie = P \lambda_i^2$. Note that $\xlie$ ranges from $r_1 $
to 0 and we have $r_1 > \lambda_i^1$. Also note that
\begin{equation}
 \max\limits_{\xlie} (\xlie)^{\alpha}(\mu - \xlie - P \xB) > \max\limits_{\xlie} (\xlie)^{\alpha}(\mu - \xlie -
 \xB) \nonumber
\end{equation}
since $P \xB < \xB$. Thus, $U_{i, max}^2 > U_{i, max}^1$.
\end{IEEEproof}

We show in Lemma \ref{lemma:3} that given $\xB$, $U_i(\lambda_i,
\lambda_{-i})$ achieves the maximal point when $P < 1$ under the
condition that $\xB + \frac{(\mu - \xB)\xa}{\xa + 1} \leq r_1$. When
$\xB + \frac{(\mu - \xB)\xa}{\xa + 1} > r_1$, there is only one
maximal point for $U_i(\lambda_i, \lambda_{-i})$. When $\xB +
\frac{(\mu - \xB)\xa}{\xa + 1} = r_1$, $U_{i, max}^1$ and $U_{i,
max}^2$ will overlap, and since $r_1 > \xopt$, this case could not
result in a NE and thus we do not consider this case.

Lemma \ref{lemma:1}, Lemma \ref{lemma:2} and Lemma \ref{lemma:3}
shows that given $\xB$, $U_i(\lambda_i, \lambda_{-i})$ achieves the
maximal point when $P < 1$. Then we will prove the uniqueness of NE
based on the expression $P(\sum \xli) = A (\xli + \xB) + D$.

\begin{IEEEproof}[Proof of Theorem \ref{claim:c6}]

Suppose the $G_p$ has more than one NE. Note that the game is
symmetric, so there must exist one NE $\lambda = \{\lambda_1,
\lambda_2, \dots, \lambda_m\}$, such that $\exists \; i,j, \lambda_i
\neq \lambda_j$. This means there exists a NE $\lambda = \{\xli,
\xB\}$ and a constant $c$ such that $\lambda = \{\xli+c, \xB-c\}$ is
also a NE.

We insert $\lambda = \{\xli, \xB\}$ into (\ref{equ:12}) and we get:
\begin{equation}
\begin{split}
&[A(\xli + \xB)+D]^2 [(\xa + 1) (\xopt - \xBe - \xlie) + \xBe] \nonumber\\
&= A \xlie (\xa +1)(\xBe + \xlie - \xopt ) \nonumber
 \end{split}
\end{equation}
\begin{equation}
\label{equ:18}
\begin{split}
\Longrightarrow \;&\frac{(\xa + 1) (\xopt - \xBe - \xlie) + \xBe}{\xlie}\\
&=\frac{1}{[A(\xli + \xB)+D]^2} A (\xa +1)(\xBe + \xlie - \xopt )
 \end{split}
\end{equation}

We also insert $\lambda = \{\xli+c, \xB-c\}$ into (\ref{equ:12}) and
denote $\tilde{c} = \frac{c}{A(\xli + \tilde{c} + \xB -
\tilde{c})+D} = \frac{c}{A(\xli + \xB)+D}$. We have:

\begin{equation}
 \begin{split}
&[A(\xli + \tilde{c} + \xB - \tilde{c})+D]^2 \\
&\qquad\qquad [(\xa + 1) (\xopt - \xBe
-\tilde{c}- \xlie + \tilde{c}) + \xBe-\tilde{c}] \nonumber\\
&= A (\xlie + \tilde{c}) (\xa +1)(\xBe - \tilde{c} + \xlie +
\tilde{c} - \xopt
 ) \nonumber
\end{split}
\end{equation}
\begin{equation}
\label{equ:19}
\begin{split}
 \Longrightarrow \;& \frac{(\xa + 1) (\xopt - \xBe -
\xlie) + \xBe - \tilde{c}}{\xlie + \tilde{c}} \\
& = \frac{1}{[A(\xli + \xB)+D]^2} A (\xa +1)(\xBe + \xlie - \xopt )
\end{split}
\end{equation}

Note that the right-hand side of (\ref{equ:18}) and (\ref{equ:19})
are the same and thus the left-hand side of (\ref{equ:18}) and
(\ref{equ:19}) are equal to each other. So,

\begin{equation}
\begin{split}
&  \frac{(\xa + 1) (\xopt - \xBe - \xlie) + \xBe}{\xlie}\\
& =\; \frac{(\xa + 1) (\xopt - \xBe - \xlie) + \xBe -
\tilde{c}}{\xlie + \tilde{c}}\\\nonumber
  \Longrightarrow \;& (\xlie + \tilde{c})(\xa + 1) (\xopt - \xBe - \xlie) + (\xlie + \tilde{c})
  \xBe \\
   & =\; \xlie (\xa + 1) (\xopt - \xBe - \xlie) + \xlie (\xBe -
 \tilde{c})\\
    \end{split}
\end{equation}
\begin{equation}
\begin{split}
   \Longrightarrow \;& \tilde{c}(\xa + 1)(\xopt - \xBe - \xlie) + \tilde{c} (\xBe + \xlie) = 0\\
   \Longrightarrow \;& \tilde{c}[(\xa + 1)\xopt - \xa(\xBe + \xlie)] = 0\\
   \Longrightarrow \;& \tilde{c}[\xa \mu - \xa(\xBe + \xlie)] = 0\\
   \Longrightarrow \;& \tilde{c}\xa [\mu - (\xBe + \xlie)] = 0\\
   \Longrightarrow \;& \frac{c \xa [\mu - (\xBe + \xlie)]}{A(\xli +
  \xB)+D} = 0 . \nonumber
  \end{split}
\end{equation}
Note that $\mu - (\xBe + \xlie) > 0$. This implies that $c = 0$ and
therefore Theorem \ref{claim:c6} holds.
\end{IEEEproof}

\section{Best Response Dynamics and Convergence}\label{sec:convergence}

In this section, we show that the best response dynamic
\cite{MacKenzie:2006}, a simple learning mechanism, will lead the
queuing game to converge to the pure Nash equilibrium.

Best response dynamic is a straightforward updating rule which
proceeds as follows: whenever player $i$ has an opportunity to
revise her strategy, she will choose the best response to the
actions of all the other players in the previous round.
Mathematically, for a game $G = (N, \{\mathcal{A}_i\}, \{U_i\})$,
let $a_i^t$ denotes the action of player $i$ in iteration $t$,
\begin{equation}
  a_i^t = \arg\max_{a_i' \in \mathcal{A}_i} U_i(a_i', a_{-i}^{t-1}).
\end{equation}

In general, the best response dynamic is not guaranteed to converge.
However, if the process does converge, it is guaranteed to converge
to a NE. Now, we want to investigate the convergence of our proposed
M/M/1 Game with the packet dropping scheme, denoted as $G_p$ in the
previous section.

\theorem\label{theorem:10} Best response dynamic will converge to
the unique NE for the M/M/1 Game with the proposed packet dropping
scheme.

\begin{IEEEproof}

There is an important result about the convergence for ordinal
potential game as shown in Theorem $21$ in \cite{MacKenzie:2006}: if
$G$ is an ordinal potential game with a compact action space and a
continuous potential function, then the best response dynamic will
(almost surely) either converge to a NE or every limit point of the
sequence will be a NE.

We have showed in Theorem \ref{claim:potential} that $G_p$ is an
ordinal potential game, and although the original definition of the
game does not have a compact action space, the equivalent
modification as shown in (\ref{eqn:compactStrategy}) has a compact
action space. We can also see that the potential function is
continuous. We have also proved in Theorem \ref{claim:c6} that there
is a unique NE. Thus Theorem \ref{theorem:10} holds.
\end{IEEEproof}

\begin{figure}[h]
\centering
\includegraphics[width=0.37\textwidth]{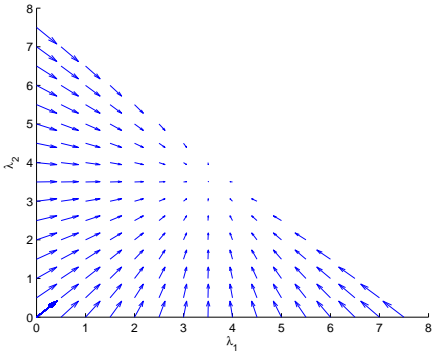}
\caption{Quiver plot for a two user example with $\mu = 10$, $\xa =
 2$, $r_1 = 7.0321$ and $r_2 = 7.8222$. The vector length are scaled
 to $\frac{1}{14}$ of the original length.} \label{fig:7}
\end{figure}

Figure \ref{fig:7} is an illustration of Theorem \ref{theorem:10} by
the quiver plot. In Figure \ref{fig:7}, on the lower triangle of a
grid (i.e., the feasible operating domain), we plot vector summation
for a two-user rate control queuing game with $\mu = 10$, $\xa = 2$,
$r_1 = 7.0321$ and $r_2 = 7.8222$. At each point, the vectors'
projections on $\lambda_1$ and $\lambda_2$ represent the best
response for the corresponding users in next iteration. To make the
plot neat, the length of each vector is scaled to $\frac{1}{14}$ of
the original length. The figure shows that at each point, the
players in the best response dynamic move towards the equilibrium
point. The length of the best response vectors are proportional to
the distance from the equilibrium point. At the equilibrium point,
the step size of the movement in the next iteration tends to zero,
which implies the convergence of the best response dynamic.


\section{Impact of Arrival Rate Estimation \label{sec:estimation}}

For a real system implementation, the server needs to estimate the
total arrival rate from users in order to apply the incentive packet
dropping scheme. The packets arrive randomly over time, so there
will be a difference between the estimated total rate and the
average total rate. This inaccuracy will cause a loss in the PoA.
While applying the packets dropping scheme, we note that the closer
to 1 the desired PoA is, the steeper (on the linear part) the packet
dropping scheme is, and therefore, the greater the impact of
estimation inaccuracy will be. So as the desired PoA approaches 1,
on the one hand the PoA of the real system should increase due to
the implementation of the incentive scheme, but on the other hand,
the sensitivity to estimation error will reduce the gain in PoA.
Therefore, the achieved PoA in practice may not be arbitrarily close
to 1.

\begin{figure*}[h]
    \centering
    \subfigure[$\mu = 500$.]
    {
        \includegraphics[width=0.31\textwidth]{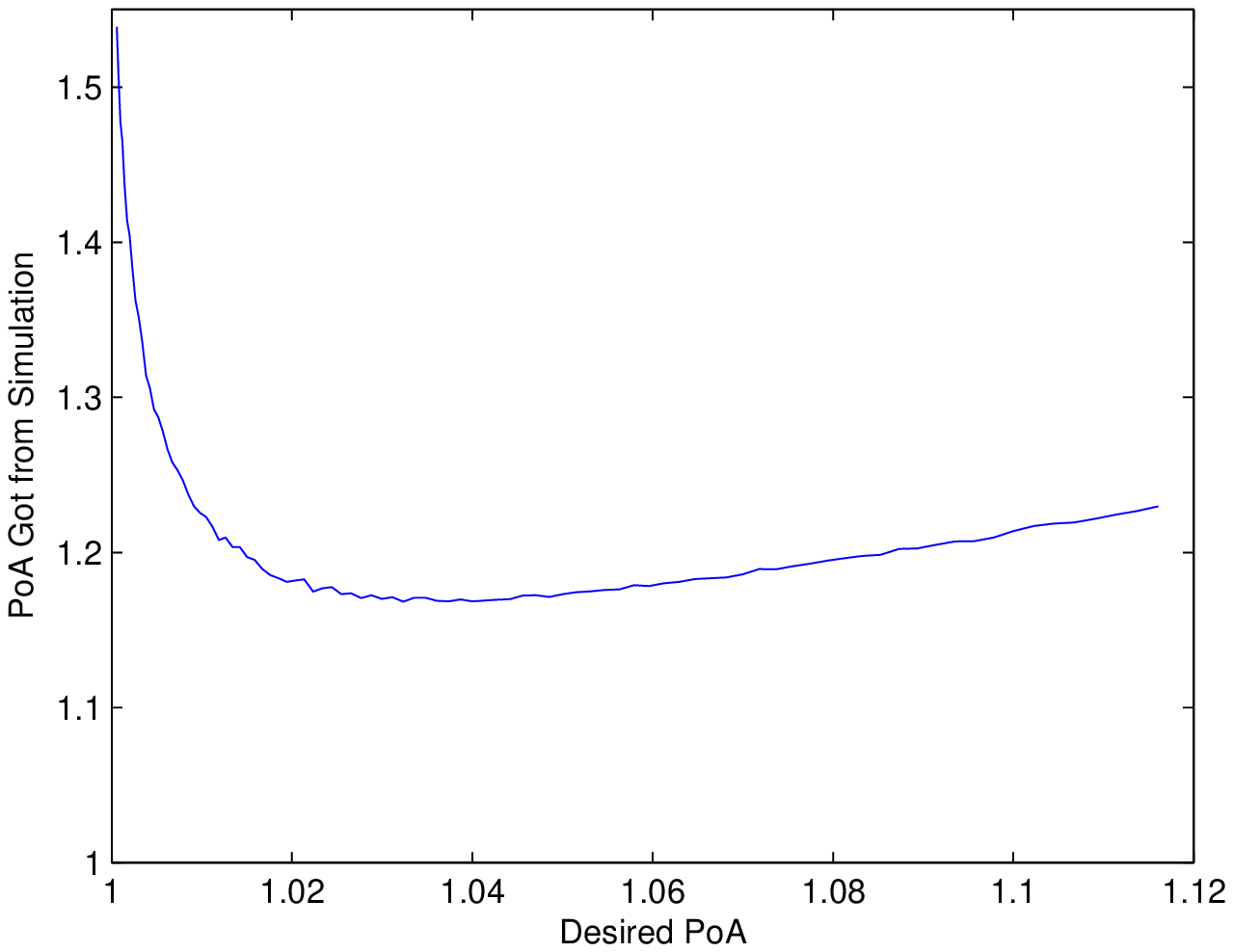}
        \label{fig:a1}
    }
    \subfigure[$\mu = 5000$.]
    {
        \includegraphics[width=0.31\textwidth]{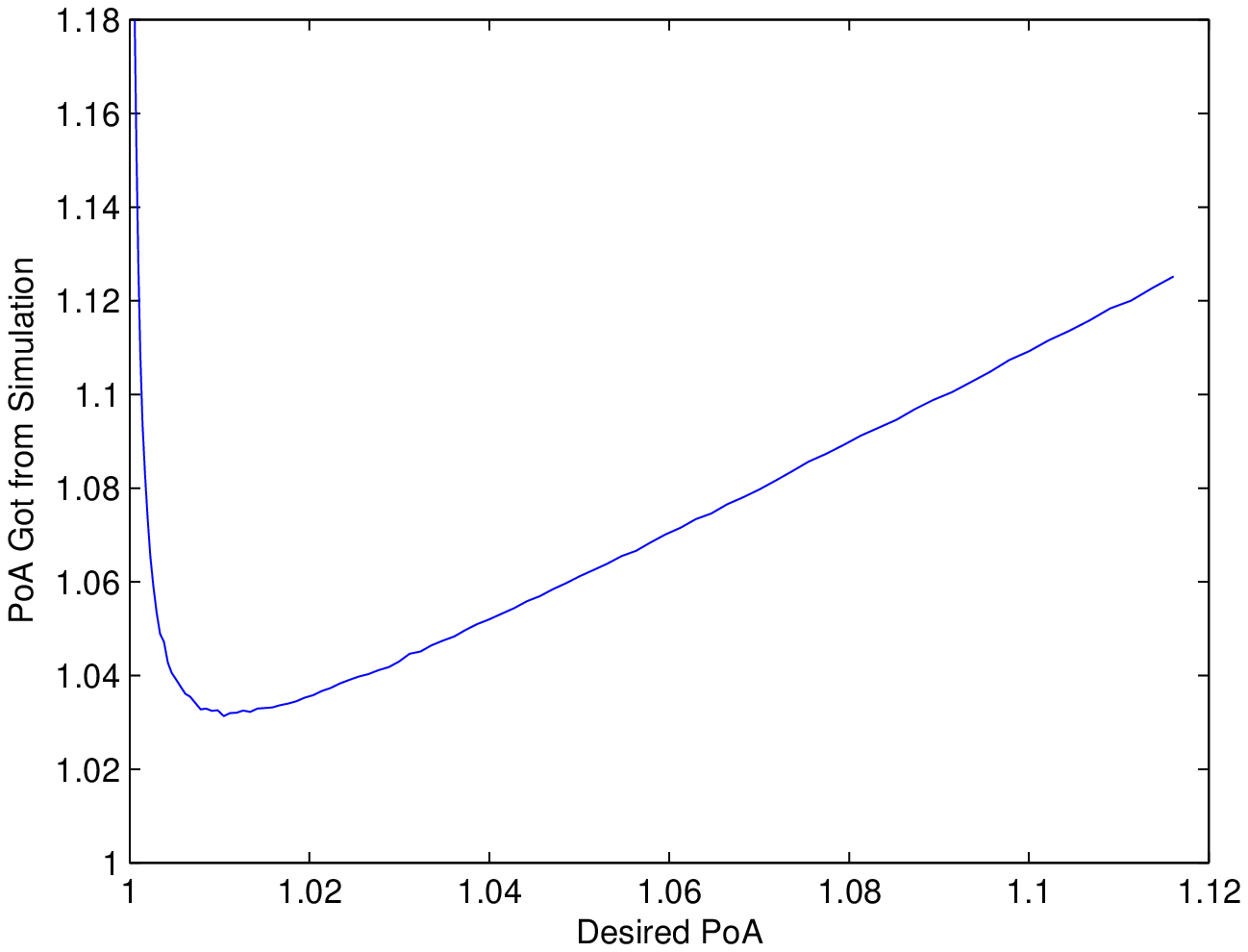}
        \label{fig:a2}
    }
    \subfigure[$\mu = 50000$.]
    {
        \includegraphics[width=0.31\textwidth]{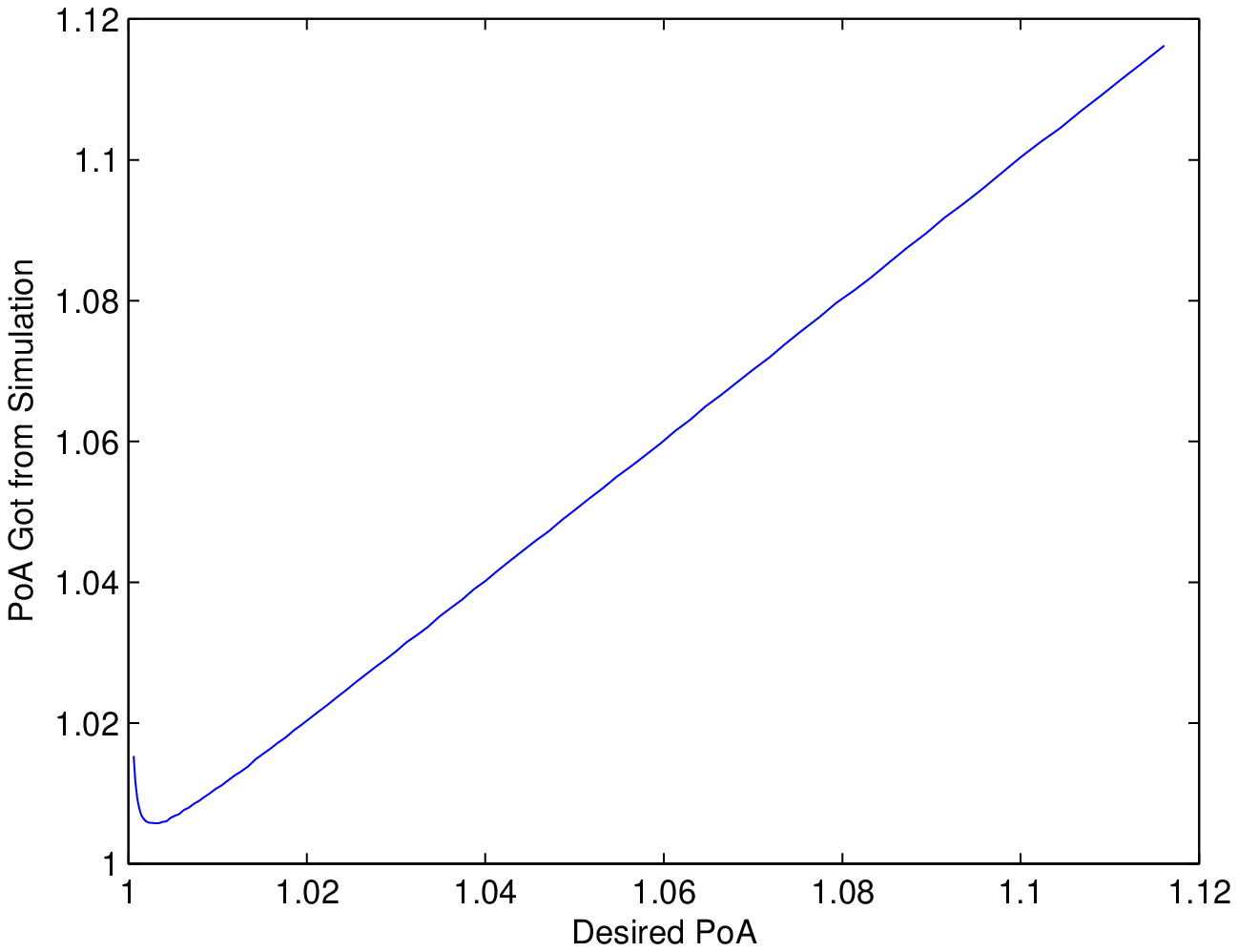}
        \label{fig:a3}
    }
    \caption{Simulation results of PoA under different service rates of a 3-user system with $\alpha = 2$ (sum-utility definition).}
    \label{fig:8}
\end{figure*}

\begin{figure*}[h]
    \centering
    \subfigure[$\mu = 500$.]
    {
        \includegraphics[width=0.31\textwidth]{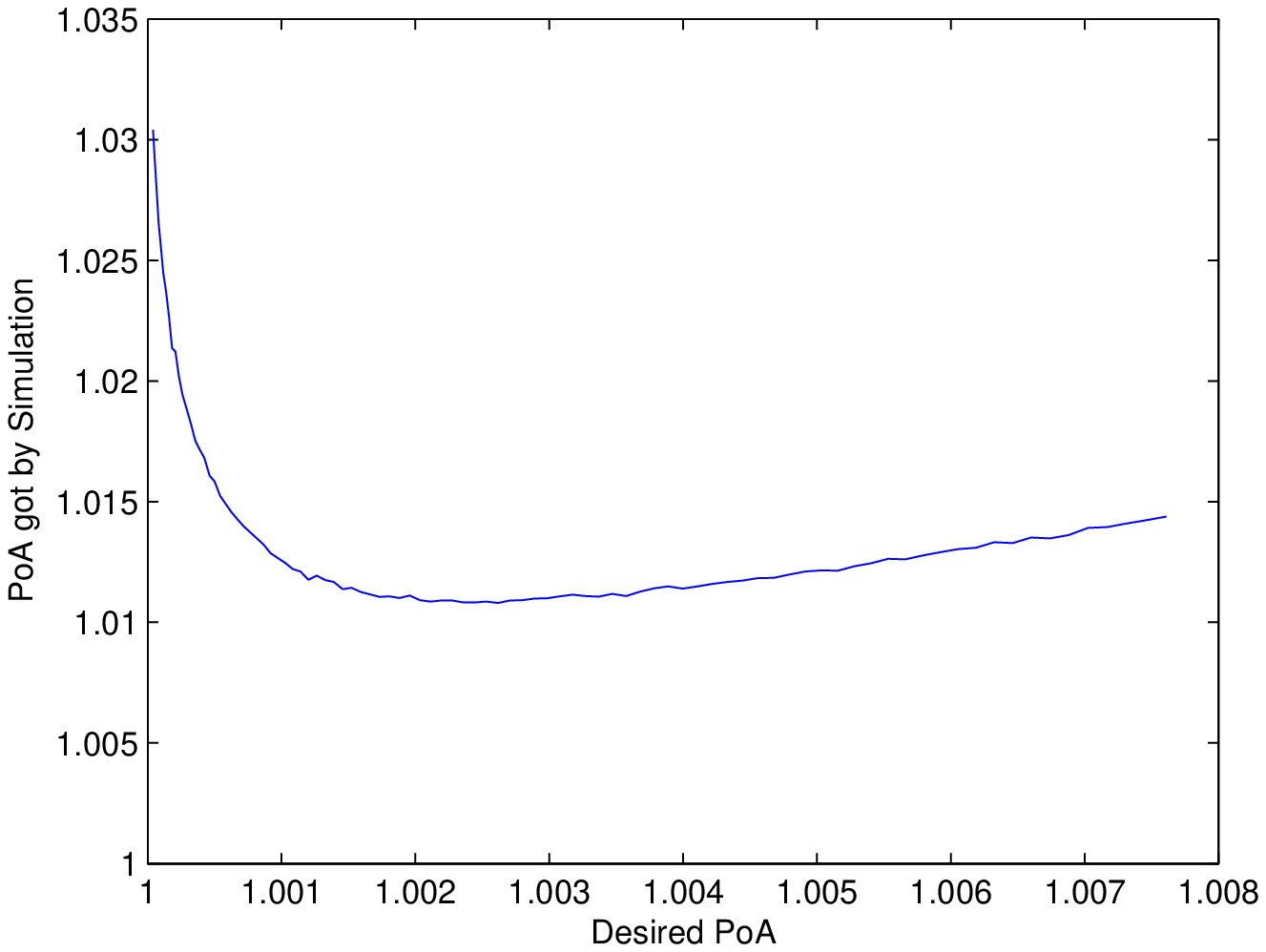}
        \label{fig:b1}
    }
    \subfigure[$\mu = 5000$.]
    {
        \includegraphics[width=0.31\textwidth]{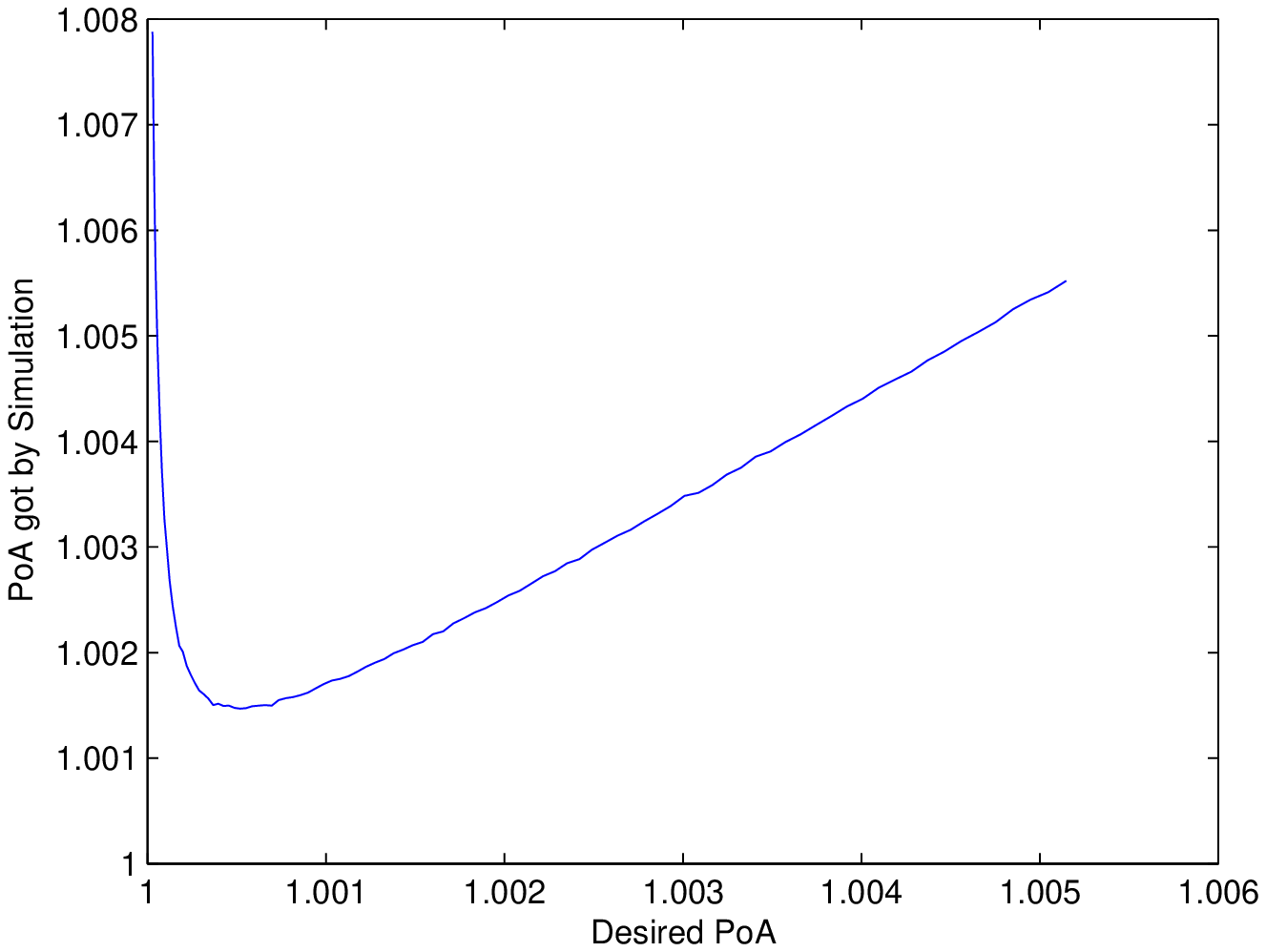}
        \label{fig:b2}
    }
    \subfigure[$\mu = 50000$.]
    {
        \includegraphics[width=0.31\textwidth]{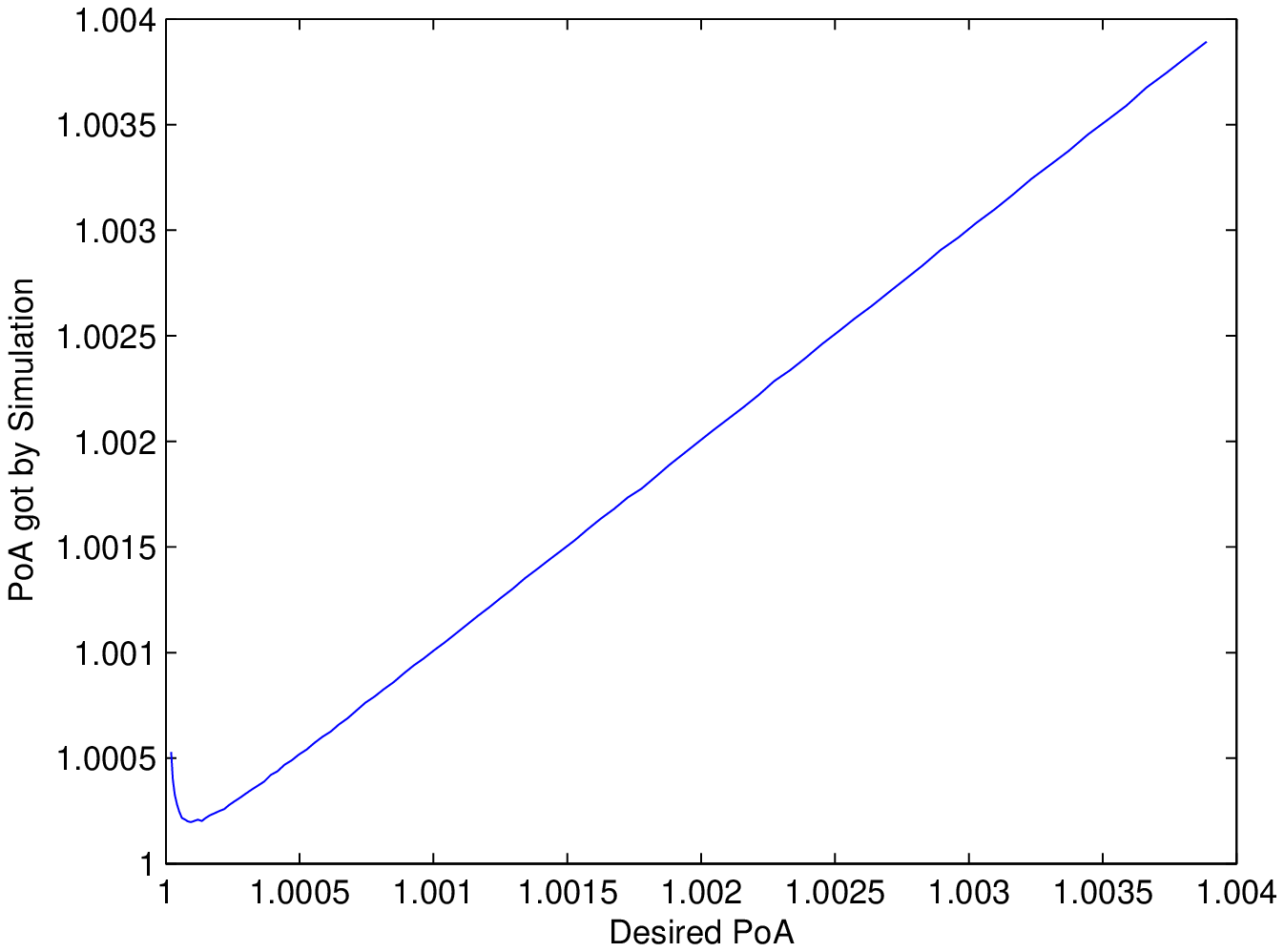}
        \label{fig:b3}
    }
    \caption{Simulation results of PoA under different service rates of a 3-user system with $\alpha = 2$ (sum-log-utility definition).}
    \label{fig:9}
\end{figure*}

\begin{figure*}[p]
    \centering
    \subfigure[The estimated total rate got based on the instantaneous arrival rate.]
    {
        \includegraphics[width=0.31\textwidth]{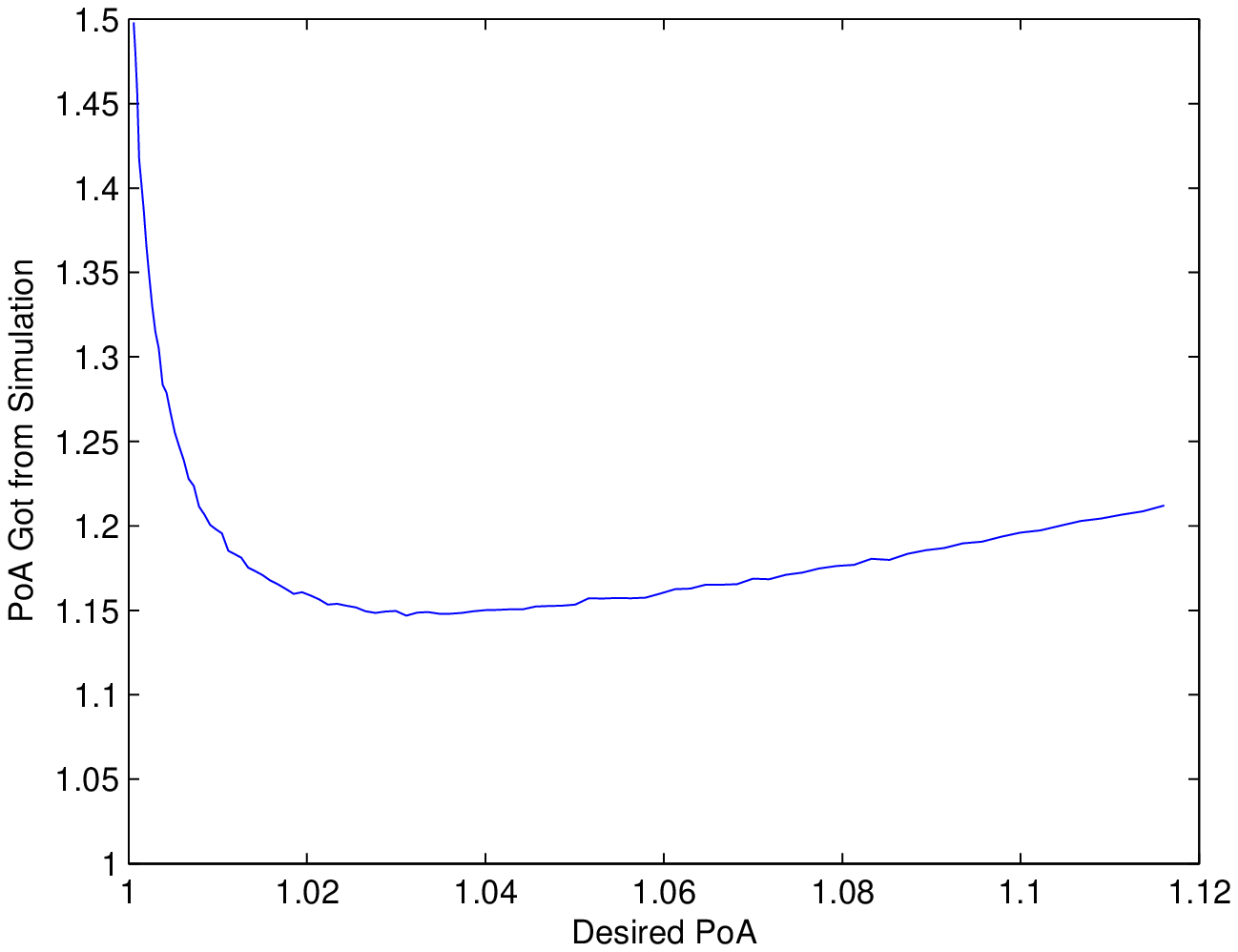}
        \label{fig:c1}
    }
    \subfigure[The estimated total rate got by averaging the continuous arrival rates in 10 slots.]
    {
        \includegraphics[width=0.31\textwidth]{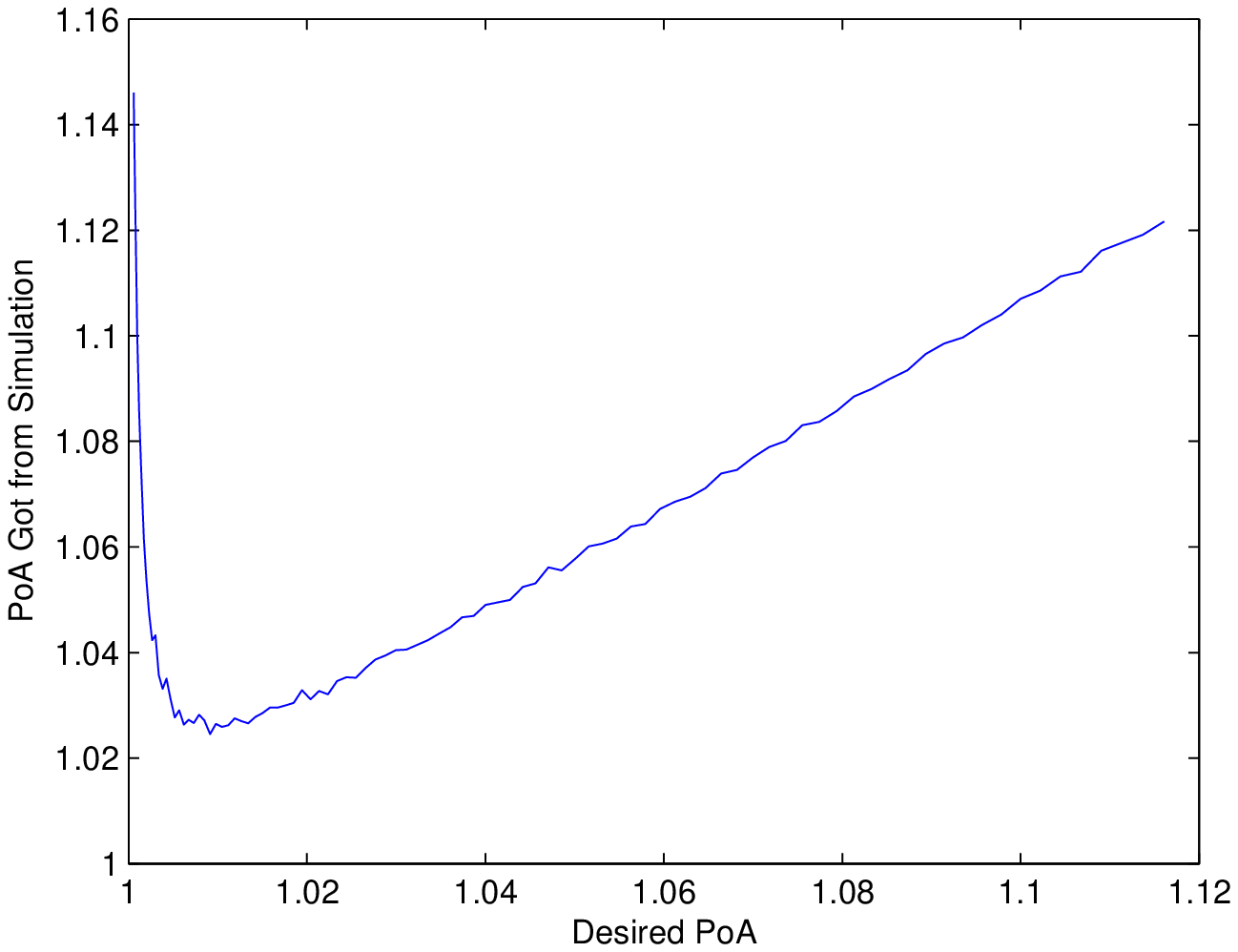}
        \label{fig:c2}
    }
    \subfigure[The estimated total rate got by averaging the continuous arrival rates in 100 slots.]
    {
        \includegraphics[width=0.31\textwidth]{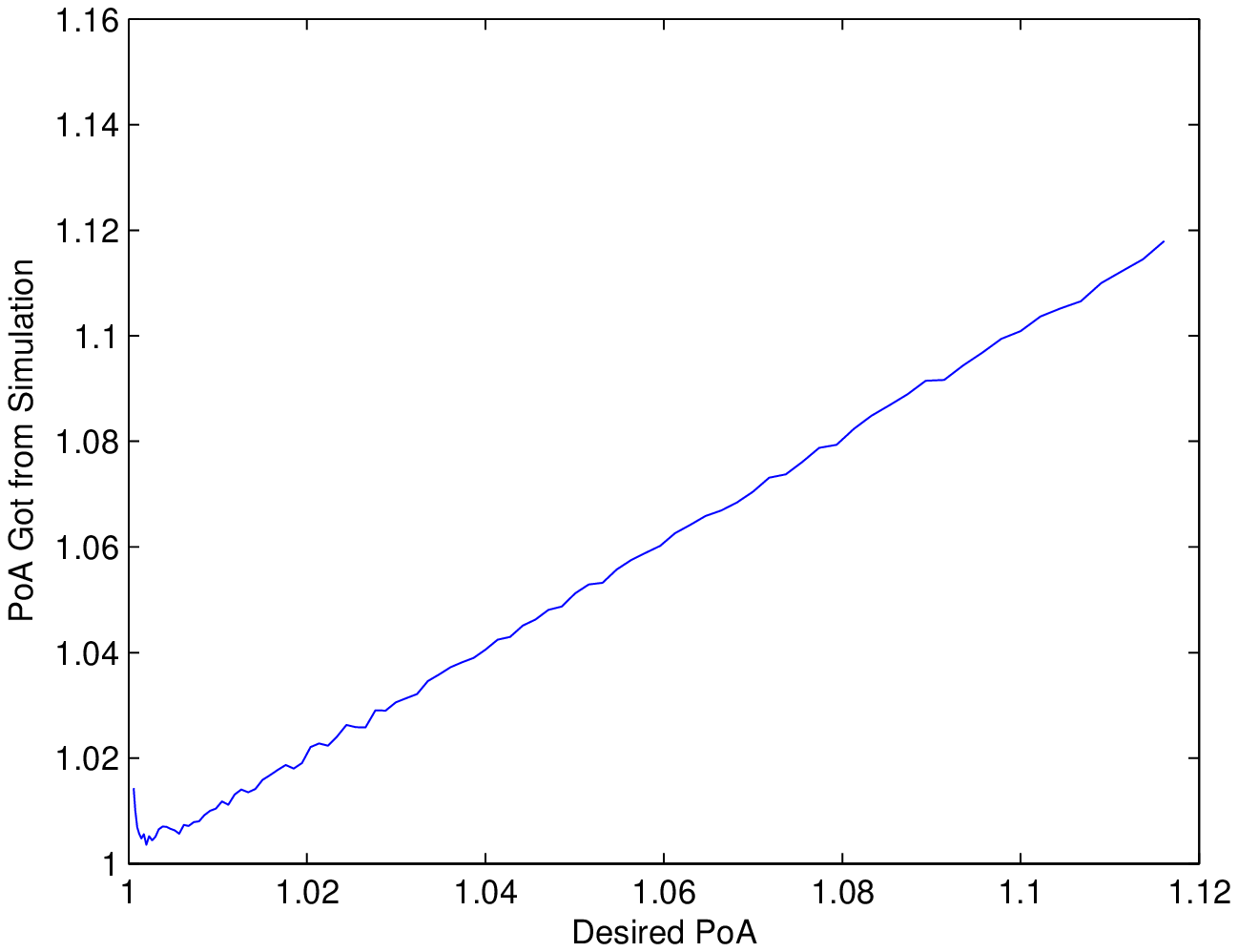}
        \label{fig:c3}
    }
    \caption{Impact of estimation length on PoA of a 3-user system with $\mu = 600$, $\alpha = 2$ (sum-utility definition).}
    \label{fig:10}
\end{figure*}

\begin{figure*}[p]
    \centering
    \subfigure[The estimated total rate got based on the instantaneous arrival rate.]
    {
        \includegraphics[width=0.31\textwidth]{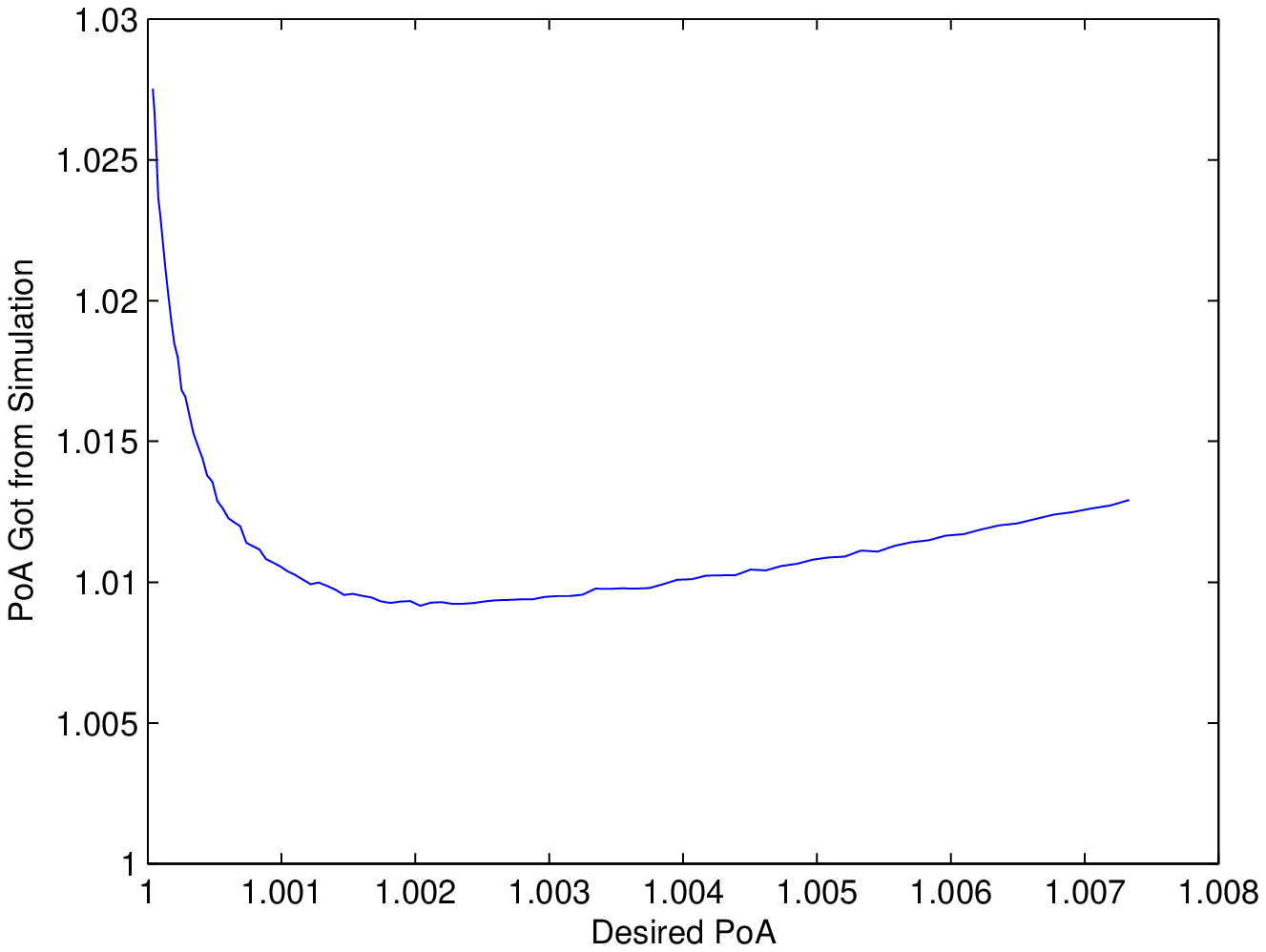}
        \label{fig:d1}
    }
    \subfigure[The estimated total rate got by averaging the continuous arrival rates in 10 slots.]
    {
        \includegraphics[width=0.31\textwidth]{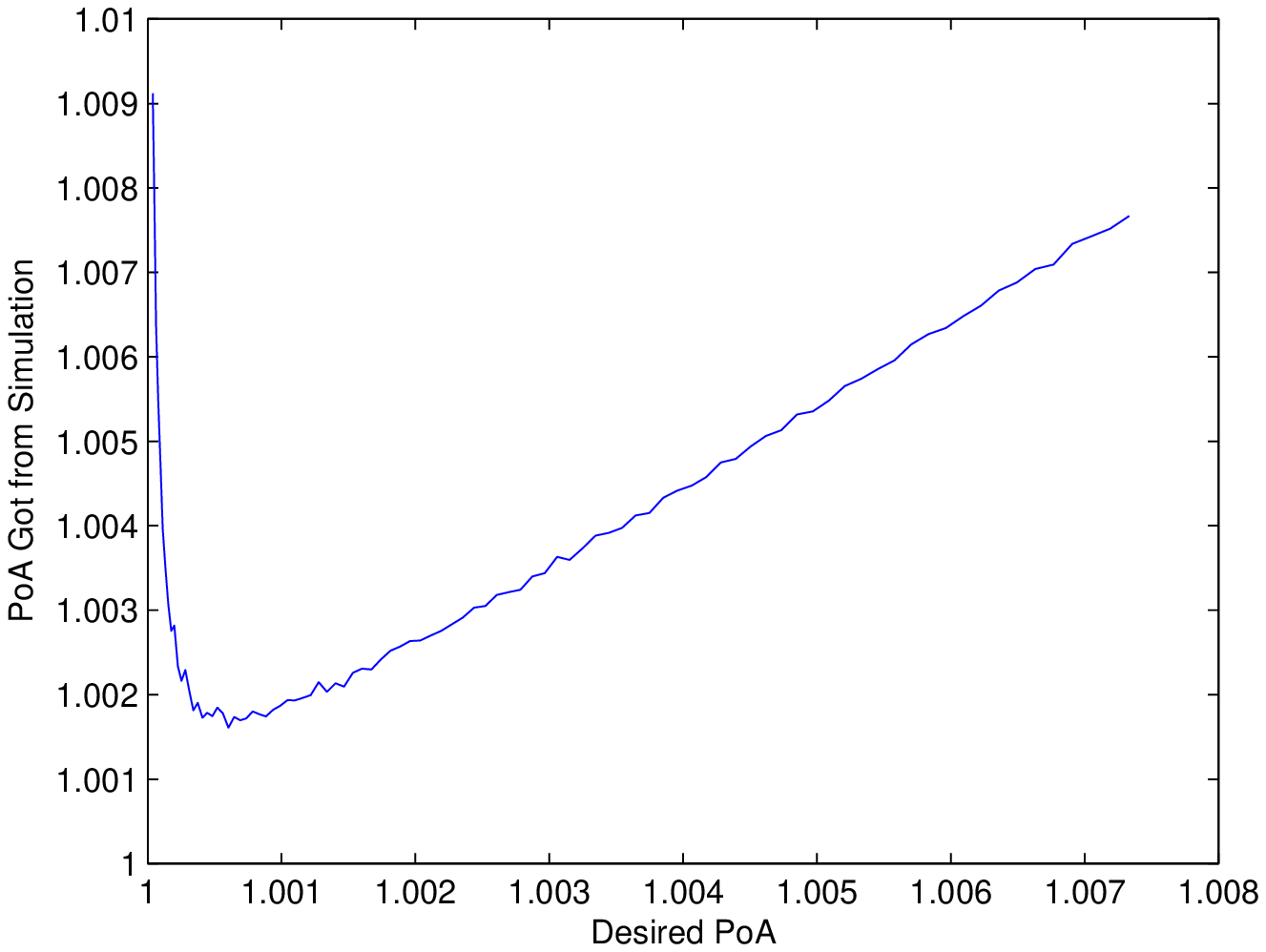}
        \label{fig:d2}
    }
    \subfigure[The estimated total rate got by averaging the continuous arrival rates in 100 slots.]
    {
        \includegraphics[width=0.31\textwidth]{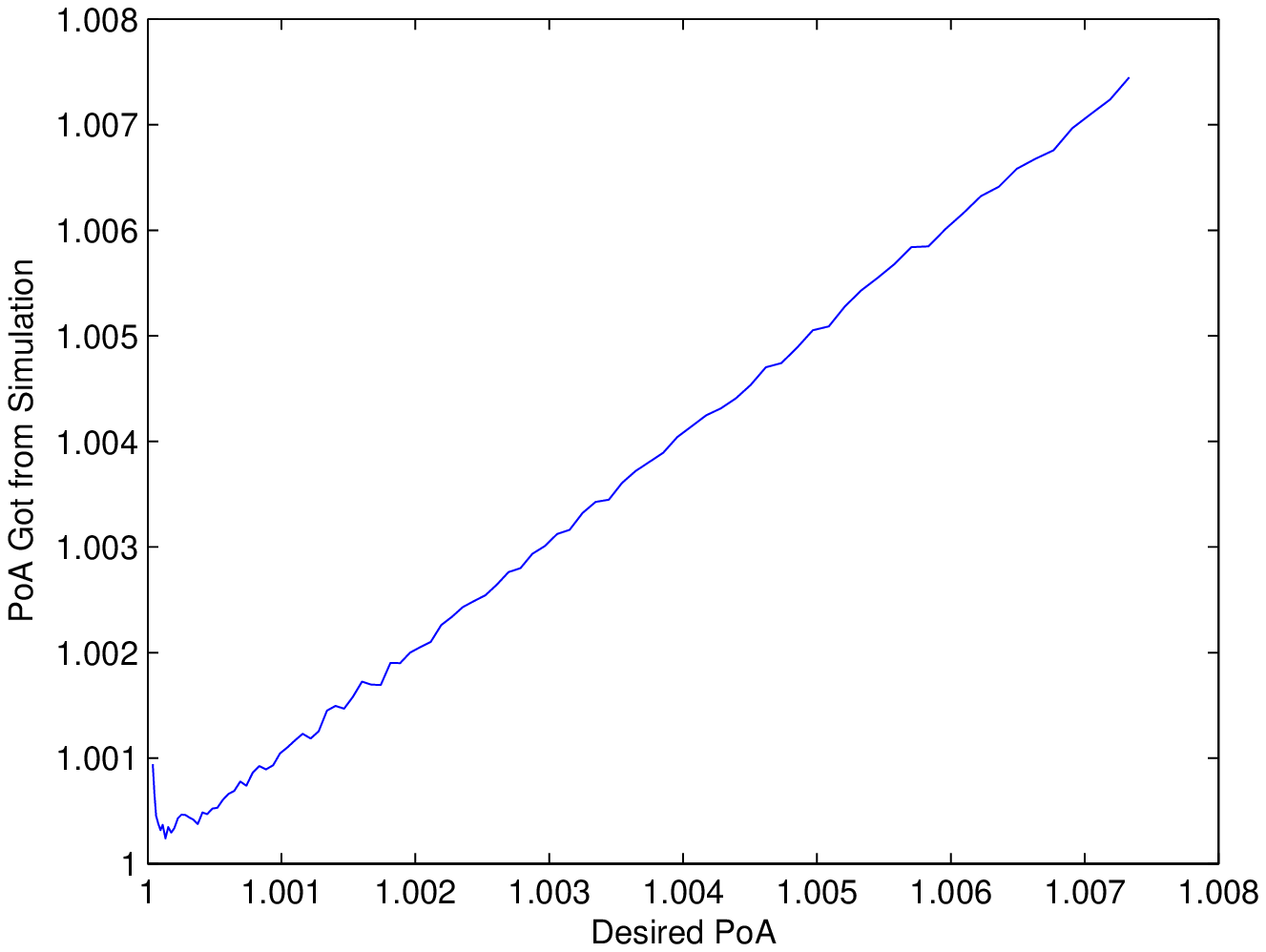}
        \label{fig:d3}
    }
    \caption{Impact of estimation length on PoA of a 3-user system with $\mu = 600$, $\alpha = 2$ (sum-log-utility  definition).}
    \label{fig:11}
\end{figure*}

We show this fact in Figure~\ref{fig:8} to~\ref{fig:11} where we
show simulation results from a 3-user queue with $\alpha = 2$. In
these simulations, we discretize time into slots. The server
estimates the mean arrival rate from the previous time slot and
applies the packet dropping function corresponding to a desired PoA
to all packets in the current slot (the users contribute arrivals at
a constant rate that corresponds to the equilibrium input for this
desired PoA). The running time for all the simulations is $10^5$
time slots. Figure~\ref{fig:8} and \ref{fig:9} show the simulation
results of PoA under different service rates for both the
sum-utility definition and sum-log-utility definition with the
instantaneous arrival rate to the server as the estimated arrival
rate. We can see that as the service rate varies from 500 to 5000
packets per time slot, the optimal point of PoA (i.e. the lowest
achievable PoA) is getting closer to 1 because the estimation
inaccuracy decreases. Also, the empirically achieved PoA is getting
closer to the desired PoA.

Note that when the service rate is low, the achievable PoA we get
could be very bad as shown in figure~\ref{fig:a1} and \ref{fig:b1}.
In these cases, using more history data/longer estimation lengths
will help to increase the estimation accuracy and improve PoA. The
comparison results under different estimation lengths for $\mu =
600$ packets per time slot are shown in figure~\ref{fig:10} and
\ref{fig:11}. These simulation results illustrate a tradeoff between
the optimal PoA and the overhead in computing and storage: while
estimating with more history data will increase the estimation
accuracy and therefore increase PoA, it increases the overhead in
terms of computing and storage.

%

\section{Conclusion}\label{sec:conclusion}
In this paper, we have designed a novel incentive mechanism for
M/M/1 queueing games with throughput-delay tradeoffs. Because the
original game yields an inefficient Nash equilibrium, we propose to
implement a linear packet dropping mechanism at the router. We show
how the parameters of this mechanism can be optimized to ensure
system efficiency that is arbitrarily close to the social welfare
solution. Further, we prove that the proposed modification has a
unique NE, and that the simple best response dynamics converges to
this solution. Future work could consider extensions of this work to
consider non-homogeneous users, other queuing models beyond the
M/M/1 model, more complex arrangements of multiple routers in a
network, as well as other system issues that may arise in practical
implementations.

\section*{Acknowledgment}

We would like to thank Professor Rahul Jain at University of
Southern California for his valuable comments.


\appendices
\section{PROOF OF THEOREM \ref{claim:1}}\label{appendix:proof01}

\begin{IEEEproof}

\begin{eqnarray}
  && \sum\limits_{i = 1}^m \log \left[ \lambda_i^{\alpha} (\mu - \sum\limits_{i = 1}^m \lambda_i) \right] \nonumber \\
  =\; &&\xa \log (\prod \xli) + m \log(\mu - \lambda)\nonumber\\
  \leq && \xa \log (\frac{ \sum \xli }{ m })^m + m \log (\mu -
 \lambda) \label{eqn:35}\\
  =\; && m \log ( \frac{\lambda}{m})^\xa (\mu - \lambda) ) \nonumber
\end{eqnarray}

Denote $f(\lambda) = \lambda^\xa (\mu - \lambda)$. So maximize
(\ref{equ:2}) is equivalent to maximize $f(\lambda)$. Take the
derivative of $f(\lambda)$ and let it equals 0. We get:
\begin{equation}
\frac{\partial f}{\partial \lambda} = 0 \Rightarrow \xopt =
\frac{\mu \xa}{\xa + 1}. \nonumber
\end{equation}
Note that equality holds in (\ref{eqn:35}) only when $\lambda_1 =
\lambda_2 = \dots = \lambda_m$. This implies $\xopt_i =
\frac{\xopt}{m} =\frac{\mu \xa}{m(\xa + 1)}$
\end{IEEEproof}

\section{PROOF OF THEOREM \ref{claim:c3}} \label{appendix:proof02}

\begin{IEEEproof} $(\Longleftarrow)$

Suppose $\sum \lambda_i' = \lambda^*$. $\forall i$,  let
$\frac{\partial U(\lambda_i, \lambda_{-i}')}{\partial \lambda_i}=
0$. We get the optimal point $\lambda_i^{**} = \frac{(\mu - \sum_{j
\neq i} \lambda_j')\alpha_i}{\alpha_i+1}$.

Note that
\begin{equation}
\begin{array}{r@{\; \;}l}
 \lambda_i' & = \lambda^* - \sum_{j \neq i} \lambda_j' = \frac{\alpha \mu}{\alpha + 1} - \sum_{j \neq i} \lambda_j' \\
 & = \frac{(\mu - \sum_{j \neq i} \lambda_j')\alpha_i - \sum_{j \neq i} \lambda_j'}{\alpha_i+1} < \frac{(\mu - \sum_{j \neq i}
 \lambda_j')\alpha_i}{\alpha_i+1} = \lambda_i^{**} . \nonumber
\end{array}
\end{equation}

Also, $\forall \lambda_i < \lambda_i^{**}, \frac{\partial
U(\lambda_i, \lambda_{-i}')}{\partial \lambda_i}> 0$, which means
$U(\lambda_i, \lambda_{-i}')$ increases monotonically with respect
to $0 \leq \lambda_i < \lambda_i^{**}$, so $U(\lambda_i',
\lambda_{-i}') > U(\lambda_i, \lambda_{-i}'), \forall \lambda_i \in
[0,\lambda_i')$. Also note that $U(\lambda_i, \lambda_{-i}') = 0,
\forall \lambda_i \in (\lambda_i^*, \mu - \sum_{j \neq i}
\lambda_j'$). Hence $\lambda_i' \in B_i (\lambda_{-i}')$.

Therefore, $\lambda'$ is a N.E.

$(\Longrightarrow)$

Suppose $\lambda'$ is a N.E., $\forall i, \lambda_i' \in B_i
(\lambda_{-i}')$.

$\forall \lambda_{-i}'$, consider the following two cases:

\subsubsection{$\sum_{j \neq i} \lambda_j' \leq \lambda^*$}\

Denote $\lambda_i'' = \lambda^* - \sum_{j \neq i} \lambda_j' $. Then
$<\lambda_i'', \lambda_{-i}'>$ is a N.E., $\lambda_i'' \in B_i
(\lambda_{-i}')$. $U(\lambda_i'', \lambda_{-i}') = U(\lambda_i',
\lambda_{-i}') > 0$. So $\lambda_i' \leq \lambda^* - \sum_{j \neq i}
\lambda_j' = \lambda''$ (if not so, $U(\lambda_i', \lambda_{-i}') =
0$).

Note that $U(\lambda_i, \lambda_{-i}')$ increases monotonically with
respect to $0 \leq \lambda_i < \lambda_i''$. Therefore, $\lambda' =
\lambda''$. $\sum \lambda_i' = \lambda^*$.

\subsubsection{$\sum_{j \neq i} \lambda_j' > \lambda^*$}\

Under this case, since $\sum \lambda'>\lambda^*$, we have $B_i
(\lambda_{-i}') =0 , \forall i$. Then $\sum_{j \neq i} \lambda_j'
\geq \lambda^*$ holds for all $i$.

$\forall i, \lambda_i' < \mu - \sum_{j \neq i} \lambda_j' < \mu -
\lambda^* = \mu - \frac{\mu \alpha}{\alpha + 1} = \frac{\mu}{\alpha
+ 1}$. So
\begin{equation}\label{equ5}
 \sum_{i = 1}^m \lambda_i' < \frac{\mu m}{\alpha + 1} < \mu \Rightarrow m < \alpha +
 1.
\end{equation}
However, note that
\begin{equation}\label{equ6}
 \lambda^* = \frac{\mu \alpha}{\alpha + 1} < \sum_{j \neq i} \lambda_j' < \frac{\mu (m-1)}{\alpha +
 1} \Rightarrow m > \alpha + 1
\end{equation}

Since (\ref{equ5}) and (\ref{equ6}) contradict each other, there is
no N.E. $\lambda'$ such that $\sum_{j \neq i} \lambda_j' >
\lambda^*$.

\end{IEEEproof}




%

\end{document}